# Wave propagation in non-centrosymmetric beam-lattices with lumped masses: discrete and micropolar modelling

Andrea Bacigalupo[1] and Luigi Gambarotta[2*]

[1]IMT School for Advanced Studies, Lucca, Italy
[2]Department of Civil, Chemical and Environmental Engineering, University of Genova, Italy

**Abstract**

The in-plane acoustic behavior of non-centrosymmetric lattices having nodes endowed with mass and gyroscopic inertia and connected by massless ligaments with asymmetric elastic properties has been analysed through a discrete model and a continuum micropolar model. In the first case the propagation of harmonic waves and the dispersion functions have been obtained by the discrete Floquet-Bloch approach. It is shown that the optical branch departs from a critical point with vanishing group velocity and for the considered cases this branch is decreasing for increasing the norm of the wave vector from the long wave limit. A micropolar continuum model, useful to approximate the discrete model, has been derived through a continualization method based on a down-scaling law from a second-order Taylor expansion of the generalized macro-displacement field. It is worth noting that the second order elasticity tensor coupling curvatures and micro-couples turns out to be negative defined also in the general case of non-centrosymmetric lattice. The eigenvalue problem governing the harmonic propagation in the micropolar non-centrosymmetric continuum results in general characterized by a hermitian full matrix that is exact up to the second order in the wave vector.

Examples concerning square and equilateral triangular lattices and their acoustic properties have been analysed from both the exact Lagrangian model (within the assumed hypotheses) and the micropolar approximate model. The analysis of the influence of the model parameters on the acoustic behavior has shown that the non-centrosymmetry topology of the lattice may contribute to obtain low frequency band gaps. As occurs in the Lagrangian model, the optical dispersion branch from the micropolar model of the considered cases turns out to be decreasing for increasing the norm of the wave vector from the long wavelength limit. On the contrary, it may be easily verified that if the elastic second order positive defined tensor is assumed, derived by a first order expansion of the rotation field, the optical branch turns out to be approximated by the equivalent micropolar continuum with a lower accuracy. Finally, in consideration of the negative definiteness of the second order elastic tensor of the micropolar model, the loss of strong hyperbolicity of the equation of motion has been investigated.

* Corresponding Author, luigi.gambarotta@unige.it

## 1. Introduction

Lattice materials have proved interesting due to their acoustic properties, as already noted by the seminal book of Brillouin (1953). The periodic microstructure of these heterogeneous materials may strongly affect the elastic wave propagation (Hutchinson and Fleck, 2006) with pass and stop bands in the acoustic spectrum, namely frequency intervals over which the wave propagation can or cannot take place respectively. These effects were obtained, among the others, by Martisson and Movchan, 2003, and Phani *et al.*, 2006, who analysed the influence of the beam lattice topology on the band gap formation. Several studies involving the lattice topology on the acoustic band gap formation have been focused on both auxetic lattices, because of their dispersive properties (see Krödel *et al.*, 2014), and on periodic lattices with prescribed defects in the microstructure (Kutsenko, 2015). Further studies extended the field of investigation to chiral and anti-chiral lattices. From the fundamental paper by Prall and Lakes, 1997, a further contribution concerns the computational investigation on hexachiral and tetrachiral lattices by Spadoni *et al.*, 2009, and Tee *et al.*, 2010, respectively, who obtained stop bands in the Bloch spectrum. Chiral dynamic microstructures have been also successfully developed by Carta *et al.*, 2014, through gyroscopes embedded into the junctions of a doubly periodic lattice. Further improvements in the design of low frequency band gaps have been obtained through the insertion in the lattice structure of local resonators based on the idea by Liu *et al.*, 2000, which provide tunable band gaps (see for reference Huang *et al.*, 2009, Liu *et al.*, 2011, Lai *et al.*, 2011, and Bacigalupo and Gambarotta, 2016). By virtue of their tunable structure, optimal design has been carried out to these locally resonant systems to control the acoustic properties (see for instance Krushynska *et al.*, 2014, and Bacigalupo *et al.*, 2016a,b). It is also worth to note that locally resonant band gaps have been obtained by Wang *et al.*, 2015, by tuning the average connectivity network of the lattice without embedding additional resonating units.

Despite the wide variety of lattice topologies considered in the literature, it should be noted that all these are characterized by periodic centrosymmetric cells. Conversely, the acoustic behavior of non-centrosymmetric lattices seems not yet analysed (with the exception of Martinsson and Movchan, 2003), although this asymmetric topology could in principle contribute to useful and interesting improvements of the acoustic performances of lattice microstructures. Such circumstance motivates the present study, which is firstly focused on lattices represented as a discrete Lagrangian system of nodes endowed with mass and gyroscopic inertia, which are connected by massless ligaments with asymmetric elastic properties apt to obtain periodic non-centrosymmetric cells. For these systems, the propagation of harmonic waves and the dispersion functions are determined through the discrete Floquet-Bloch approach.

Although the discrete model presents the considerable advantage to take into account the microstructural effects in a more accurate way, in particular when the wavelength is comparable to the microstructure size, on the other hand, dynamic equivalent continuum models, which are reliable in the long-wave approximation of the lattice model, present the advantage to be based on macroscopic parameters that may allow a simpler dynamic description. In case of beam-lattices connected by massless ligaments, Suiker *et al.*, 2001, derived the equation of motion of the equivalent Cosserat continuum through a continualization method (Metrikine and Askes, 2002), namely by replacing the degrees of freedom of the neighboring cells in the discrete equation of motion with the second-order Taylor approximation of the macro-field continuous variables. With a similar continualization technique and truncating to exclude all the terms $\mathcal{O}\left(\epsilon^3\right)$ from the equation of motion, $\epsilon$ being the smallness ratio between the ligament and the structure size, Gonella and Ruzzene, 2008, obtained the set of PDEs of a homogenized equivalent classical continuum. Here, the truncation process excludes the rotational inertia of the nodes and the micro-couples from the equation of motion by virtue of the static condensation of the nodal rotation. Lombardo and Askes, 2012, applied a continualization technique with a fourth-order Taylor expansion of the macro-displacements and, after eliminating the nodal rotational dofs, obtained a continuum model with higher gradient inertia and stiffness terms. To reduce the possible numerical instability in the dispersion relations due to higher order gradients, the order of spatial derivatives in the PDE has been reduced through the Padé approximation of the differential operators. A refinement of the continualization technique applied to square lattices endowed of rotational inertia of the nodes has been proposed by Vasiliev *et al.*, 2008, 2010, 2014, to obtain multi-field continuum models, which approximates the dynamic behavior of the Lagrangian one also in the short-wavelength propagation. Liu *et al.*, 2012, have analysed hexachiral lattices and obtained the elastic moduli of the equivalent isotropic micropolar continuum and the dispersive curves showing explicitly the chiral effects. Nevertheless, they pointed out the problem already stressed by Bazant and Christensen, 1972, and Kumar and McDowell, 2004, regarding some ambiguities on the choice of the elastic moduli associated to the curvatures in the static micropolar homogenization of lattices.

This point, that seems rather central in the dynamics micropolar homogenization, has been analyzed in the second part of the paper. In this regard, it should be emphasized that several equivalent micropolar continuum models proposed in literature for the static analysis are based on down-scaling laws which approximate the microscopic displacement field through a first order expansion of the macroscopic displacement field. The overall-elastic moduli are obtained by the generalized principle of Hill-Mandel (see for example Chen *et al.*, 1998, Pradel and Sab, 1998 Onk, 2002). As pointed out by Kumar and McDowell, 2004, the constants relating the micro-couples and the curvatures so obtained differ from those derived considering a second order expansion of the generalized micro-displacement field. While in the first case these elastic moduli are positive defined, on the other side the positive definiteness is not assured in the second approach. A circumstance that does not occur in some other static homogenization techniques (Dos Reis and Ganghoffer, 2012). To better understand this issue, also in the more general case of microstructures with non-centrosymmetric cells, the equations of motion of the equivalent micropolar continuum have been derived through the continualization technique, by retaining the second-order terms in the expansion of the macro-displacement field. Hence, the elastic constants corresponding have been identified. Therefore, considering the same approximation of the micro-displacement field, the same elastic constants of the equivalent micropolar continuum have been identified also through the generalized Hill-Mandel criterion according to Kumar and McDowell, 2004, and the equations of motion have been obtained by means of the Hamilton extended principle. The equations of propagation of harmonic waves and the dispersion functions derived by the discrete Lagrangian model have been compared with those from the homogenized model having a negative defined constitutive tensor relating micro-couples to micro-curvatures and finally with the corresponding ones derived assuming the positive definite tensor, namely those obtained by the first order approach and commonly considered in standard micropolar homogenization. Finally, two lattice typologies have been analysed to show both the influence of the non-centrosymmetric topology of the periodic cell on the elastic wave propagation and Bloch spectra and the accuracy and the validity limits of the considered micropolar homogenization technique.

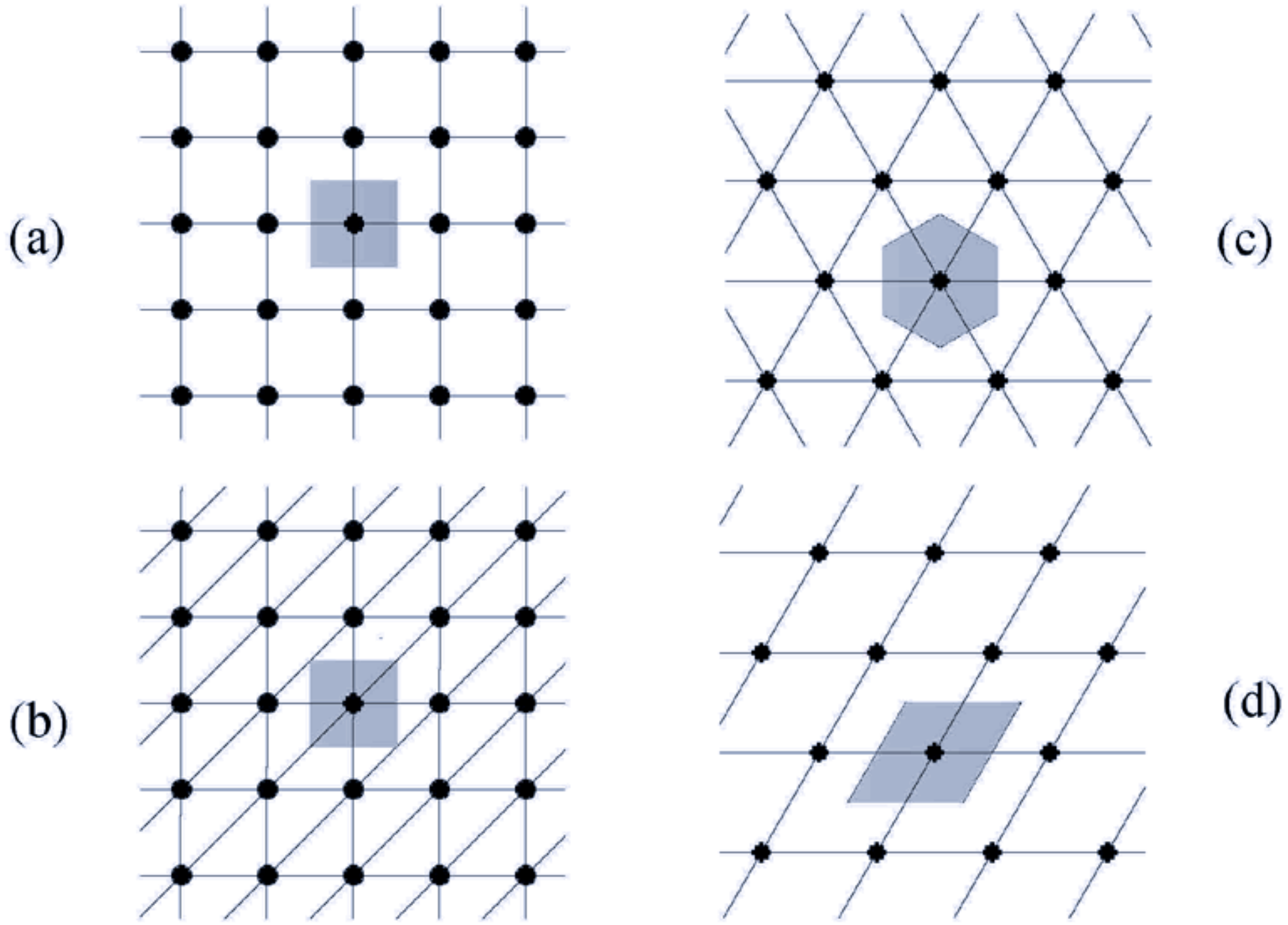

Figure 1: Examples of periodic lattices with lumped masses: (a) rectangular (*n*=4); (b) rectangular (*n*=6), (c) triangular (*n*=6), (d) rectangular deformed (*n*=4).

## 2. Lagrangian modeling of two-dimensional periodic beam-lattices

Let us consider 2-D periodic cellular solids made up of elastic slender ligaments rigidly connected at the nodes of the lattice, as shown in the examples of Figure 1. The considered topology has the property that a periodic cell may be identified with a node located at its centre and connected to the surrounding ones through an even number $n$ of ligaments. The $i$-th ligament has length $l_i$, variable section width $t_i$, unit thickness and Young modulus $E_s$. Moreover, the $i$-th ligament has a corresponding opposite ligament $j = N + i$ ($N = n/2$) with respect to the node, having the same length and Young modulus as shown in Figure 2. The section width is assumed to be variable along two opposite ligaments, with $t(A) \neq t(A')$, $A$ and $A'$ being two points on the ligament symmetric with respect to the central node (see Figure 2), and $t(B) = t(B')$ to guarantee the continuity of the section width at the cell boundary. Because of this assumption, the cell is no longer centrosymmetric. Finally, each node is assumed having mass $M$ and moment of inertia $J = Mr^2$, $r$ being the radius of gyration, while the ligament mass is ignored because the present investigation is focused mainly to the propagation of low-frequency dispersive waves.

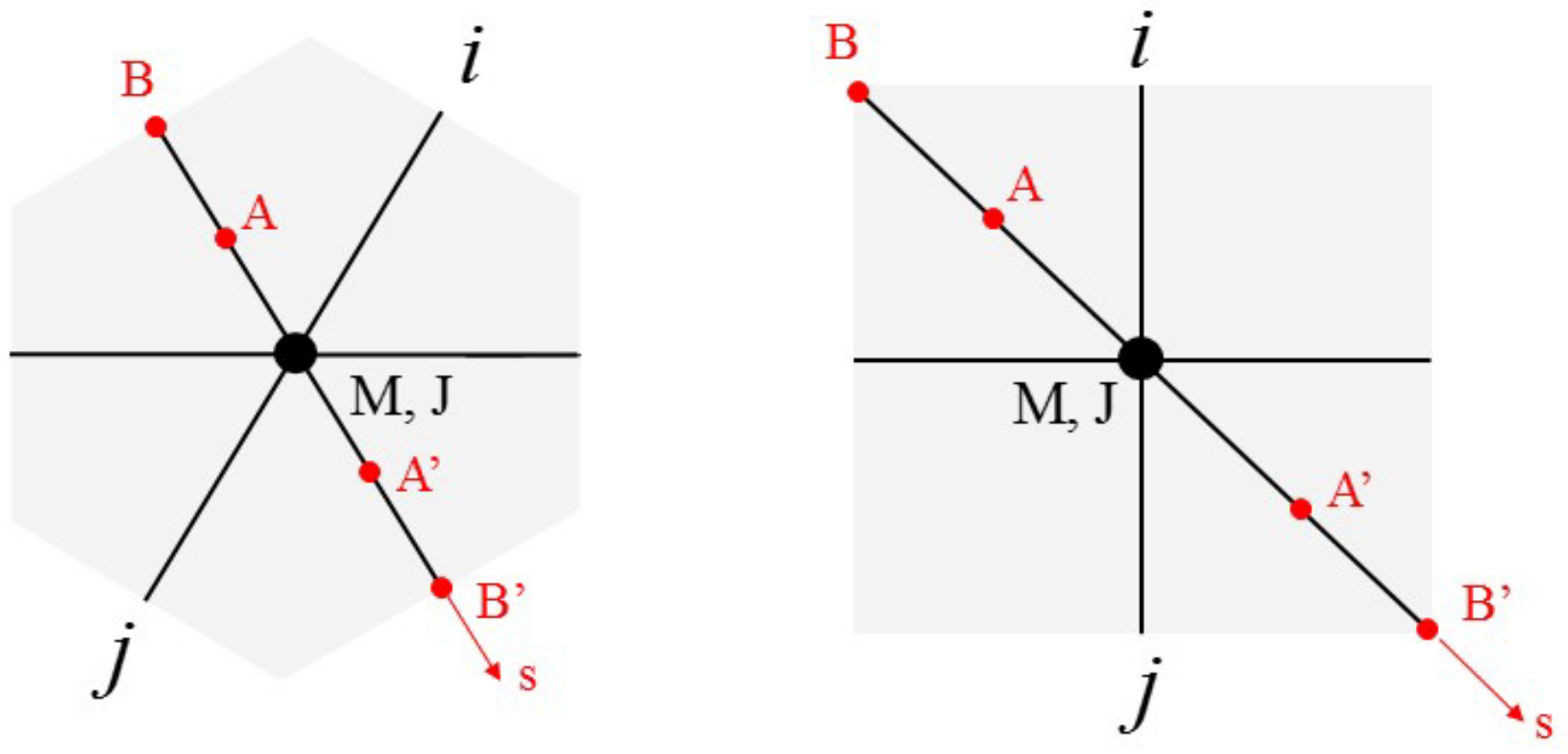

Figure 2: Symmetric points at the ligaments in the periodic cell.

Based on the previous assumption, a Lagrangian model of the beam lattice is derived with the motion of each node described by a generalized in-plane displacement, namely a translation and a rotation. The Lagrangian function $\mathcal{L} = T - \Pi$ of the lattice, $T$ being the kinetic energy and $\Pi$ the total potential elastic energy, is obtained by summing up the contributions of all its periodic cells. To derive the Euler-Lagrange equation of motion of each node, let us consider a reference periodic cell located at the origin O with nodal displacement $\mathbf{u}(t)$ and rotation $\phi(t)$ (see Figure 3). The kinetic energy of the reference cell is

$$T\left(\dot{\mathbf{u}}, \dot{\phi}\right) = \frac{1}{2} M \dot{\mathbf{u}}^2 + \frac{1}{2} J \dot{\phi}^2 \ . \tag{1}$$

To evaluate the elastic potential energy stored in the $n$ ligaments surrounding the reference cell, let us consider the $i$-th ligament connecting the central node to the $i$-th adjacent one with generalized displacement $\mathbf{u}_i$ and , $\phi_i$ (see Figure 3(a)). Here, the vector $\mathbf{x}_i = l\ \mathbf{n}_i$ connects the centre of the reference cell to the centre of the $i$-th adjacent one, being $\mathbf{n}_i$ the unit vector associated to the $i$-th ligament and $\mathbf{t}_i = \mathbf{e}_3 \times \mathbf{n}_i$ unit normal vector. The ligament extension is $\Delta_{di} = \left(\mathbf{u}_i - \mathbf{u}\right) \cdot \mathbf{n}_i$, while the transverse relative displacement between the ends of the ligament is $\Delta_{ti} = \left(\mathbf{u}_i - \mathbf{u}\right) \cdot \mathbf{t}_i$ as shown in Figure 3(b). The mean rotation of the $i$-th ligament is $\psi_i = \Delta_{ti} / l$ and the end rotation of the ligament at the central and $i$-th nodes are $\varphi = \phi - \psi_i = \phi - \frac{\Delta_{ti}}{l_i}$ and $\varphi_i = \phi_i - \psi_i = \phi_i - \frac{\Delta_{ti}}{l_i}$, respectively.

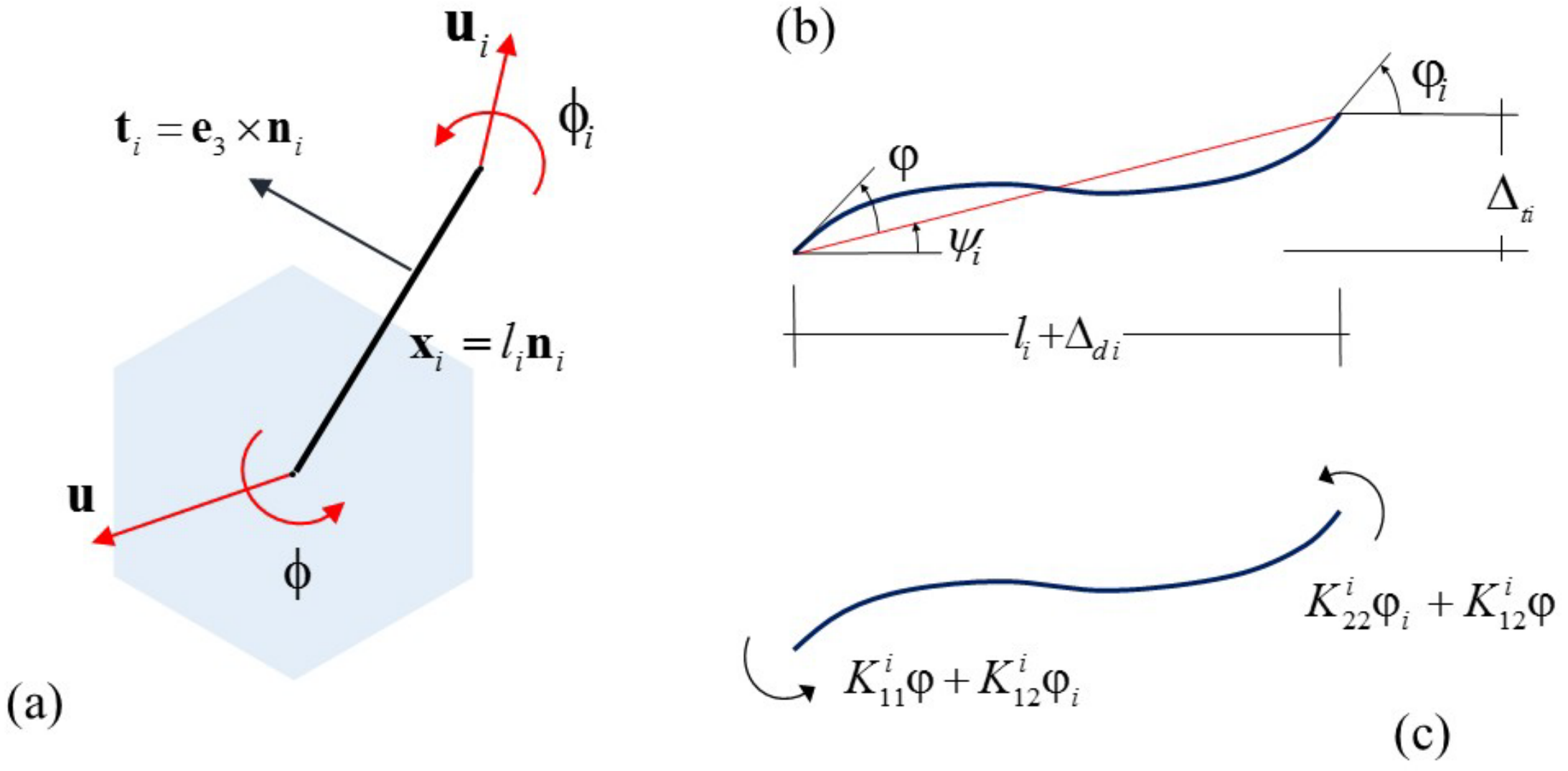

Figure 3: (a) The *i*-th ligament and the connected nodes; (b) ligament deformation; (c) ligament nodal couples.

The potential elastic energy due to axial strain $\varepsilon_i = \frac{\Delta_{di}}{l_i}$ of the $i$-th ligaments takes the form

$$\Pi_{ai}\left(\mathbf{u},\phi,\mathbf{u}_i,\phi_i\right) = \frac{1}{2}K_n^i\Delta_{di}^2 = \frac{1}{2}K_n^i\left(\mathbf{u}_i - \mathbf{u}\right)\bullet\left[\mathbf{n}_i \otimes \mathbf{n}_i\right]\left(\mathbf{u}_i - \mathbf{u}\right), \tag{2}$$

where $K_n^i$ is the axial stiffness of the $i$-th ligament. The potential elastic energy due to bending of the ligament is written as

$$\begin{aligned}\Pi_{bi}\left(\mathbf{u},\phi,\mathbf{u}_i,\phi_i\right) &= \frac{1}{2}\left(K_{11}^i\varphi^2 + 2K_{12}^i\varphi\varphi_i + K_{22}^i\varphi_i^2\right) = \\ &= \frac{1}{2}\left\{\begin{aligned}&K_{11}^i\left[\phi - \frac{1}{l_i}\mathbf{t}_i\bullet\left(\mathbf{u}_i - \mathbf{u}\right)\right]^2 + 2K_{12}^i\left[\phi - \frac{1}{l_i}\mathbf{t}_i\bullet\left(\mathbf{u}_i - \mathbf{u}\right)\right]\left[\phi_i - \frac{1}{l_i}\mathbf{t}_i\bullet\left(\mathbf{u}_i - \mathbf{u}\right)\right] + \\ &+K_{22}^i\left[\phi_i - \frac{1}{l_i}\mathbf{t}_i\bullet\left(\mathbf{u}_i - \mathbf{u}\right)\right]^2\end{aligned}\right\},\end{aligned} \tag{3}$$

$K_{11}^i$ , $K_{11}^i$, $K_{11}^i$ being the nodal stiffnesses as shown in Figure 3(c).

The Lagrangian function is obtained as the sum of the contributions of all the elementary cells

$$\mathcal{L} = \sum \left\{ T_s\left(\dot{\mathbf{u}}, \dot{\phi}\right) - \sum_{i=1}^{n} \frac{1}{2}\left[\Pi_{ai}\left(\mathbf{u}, \phi, \mathbf{u}_i, \phi_i\right) + \Pi_{bi}\left(\mathbf{u}, \phi, \mathbf{u}_i, \phi_i\right)\right] \right\} \ , \tag{4}$$

where the index denoting the cells has been omitted. The Euler-Lagrange equations of motion of the reference cell located at O are derived by the Lagrangian function (4) and depend on the generalized displacement and velocity of the node of the same cell and on the generalized displacement of the *n* surrounding nodes in the following form

$$\sum_{i=1}^{n} \left\{ \begin{array}{c} \left[ K_n^i \left(\mathbf{n}_i \otimes \mathbf{n}_i\right) + \dfrac{K_{11}^i + 2K_{12}^i + K_{22}^i}{l_i^2}\left(\mathbf{t}_i \otimes \mathbf{t}_i\right) \right] \left(\mathbf{u}_i - \mathbf{u}\right) + \\ -\dfrac{K_{11}^i + K_{12}^i}{l_i} \mathbf{t}_i \phi - \dfrac{K_{12}^i + K_{22}^i}{l_i} \mathbf{t}_i \phi_i \end{array} \right\} - \mathrm{M}\ddot{\mathbf{u}} = \mathbf{0} \, , \tag{5}$$

$$\sum_{i=1}^{n} \left[ \frac{K_{11}^i + K_{12}^i}{l_i} \mathbf{t}_i \bullet \left(\mathbf{u}_i - \mathbf{u}\right) - K_{11}^i \phi - K_{12}^i \phi_i \right] - J\ddot{\phi} = 0 \ \ . \tag{6}$$

In case of ligaments with uniform section width *t*, the equation of motion are simplified as follows:

$$E_s \sum_{i=1}^{n} \left(\frac{t}{l_i}\right) \left\{ \left[ \left(\mathbf{n}_i \otimes \mathbf{n}_i\right) + \left(\frac{t}{l_i}\right)^2 \left(\mathbf{t}_i \otimes \mathbf{t}_i\right) \right] \left(\mathbf{u}_i - \mathbf{u}\right) - \frac{l_i}{2}\left(\frac{t}{l_i}\right)^2 \mathbf{t}_i \left(\phi + \phi_i\right) \right\} - \mathrm{M}\ddot{\mathbf{u}} = \mathbf{0} \, , \tag{7}$$

$$E_s \sum_{i=1}^{n} \left(\frac{t}{l_i}\right)^3 \left\{ \frac{l_i}{2} \mathbf{t}_i \cdot \left(\mathbf{u}_i - \mathbf{u}\right) - \frac{l_i^2}{6}\left(2\phi + \phi_i\right) \right\} - \mathrm{J}\ddot{\phi} = 0 \ \ . \tag{8}$$

The propagation of a harmonic plane wave along axis $\mathbf{i}$ is investigated by substituting the following displacement field $\mathbf{U} = \hat{\mathbf{U}} \exp\left[ i\left(\mathbf{k} \cdot \mathbf{x} - \omega t\right)\right]$ in the system of three ODE’s per node, $\mathbf{k} = q\mathbf{i}$ being the wave vector and *q*, $\omega$ and $\hat{\mathbf{U}} = \left\{ \hat{\mathbf{u}}^T \quad \hat{\phi} \right\}^T = \left\{ \hat{u}_1 \quad \hat{u}_2 \quad \hat{\phi} \right\}^T$ denoting the wave number, the circular frequency and the polarization vector, respectively. A system of three linear homogeneous equations is obtained

$$\left\{\sum_{i=1}^{n}\left[K_n^i\left(\mathbf{n}_i\otimes\mathbf{n}_i\right)+\frac{K_{11}^i+2K_{12}^i+K_{22}^i}{l_i^2}\left(\mathbf{t}_i\otimes\mathbf{t}_i\right)\right]\left[1-\exp\left(i\mathbf{k}\cdot\mathbf{x}_i\right)\right]-\omega^2\mathrm{M}\mathbf{I}_2\right\}\hat{\mathbf{u}}+$$
$$+\sum_{i=1}^{n}\left\{\frac{1}{l_i}\left[K_{11}^i+K_{12}^i+\left(K_{12}^i+K_{22}^i\right)\exp\left(i\mathbf{k}\cdot\mathbf{x}_i\right)\right]\mathbf{t}_i\right\}\hat{\phi}=\mathbf{0}\ , \tag{9}$$

$$\sum_{i=1}^{n}\left\{\frac{1}{l_i}\left[K_{11}^i+K_{12}^i-\left(K_{11}^i+K_{12}^i\right)\exp\left(i\mathbf{k}\cdot\mathbf{x}_i\right)\right]\mathbf{t}_i\right\}\cdot\hat{\mathbf{u}}+$$
$$+\left\{\sum_{i=1}^{n}\left[K_{11}^i+K_{12}^i\exp\left(i\mathbf{k}\cdot\mathbf{x}_i\right)\right]\ -\omega^2 J\right\}\hat{\phi}=0\ . \tag{10}$$

If the analysis is focused on the case where to the *i*-th ligament with nodal stiffnesses $K_n^i$, $K_{11}^i$, $K_{12}^i$, $K_{22}^i$ corresponds the opposite *j*-th ligament ( $j=N+i$ ), with $\mathbf{x}_j=-\mathbf{x}_i$ and $\mathbf{t}_j=-\mathbf{t}_i$ and reversed nodal stiffnesses $K_n^j=K_n^i$, $K_{11}^j=K_{22}^i$, $K_{22}^j=K_{11}^i$, $K_{12}^j=K_{12}^i$, the system (9) and (10) takes the form of an eigenproblem ruled by a Hermitian matrix

$$\mathbf{C}_{Lag}\left(\mathbf{k},\omega\right)\hat{\mathbf{U}}=\begin{bmatrix}\mathbf{A}-\omega^2\mathrm{M}\mathbf{I}_2 & \mathbf{a}+i\mathbf{b}\\ \mathbf{a}^T-i\mathbf{b}^T & C-\omega^2\mathrm{J}\end{bmatrix}\begin{Bmatrix}\hat{\mathbf{u}}\\ \hat{\phi}\end{Bmatrix}=\mathbf{0}\,, \tag{11}$$

being the following terms defined as:

$$\mathbf{A}=2\sum_{i=1}^{N}\left[K_n^i\left(\mathbf{n}_i\otimes\mathbf{n}_i\right)+\frac{K_{11}^i+2K_{12}^i+K_{22}^i}{l_i^2}\left(\mathbf{t}_i\otimes\mathbf{t}_i\right)\right]\left[1-\cos\left(\mathbf{k}\cdot\mathbf{x}_i\right)\right]\ ,$$
$$\mathbf{a}=\sum_{i=1}^{N}\left\{\frac{K_{11}^i-K_{12}^i}{l_i}\left[1-\cos\left(\mathbf{k}\cdot\mathbf{x}_i\right)\right]\right\}\mathbf{t}_i\ ,$$
$$\mathbf{b}=\sum_{i=1}^{N}\left\{\frac{K_{11}^i+2K_{12}^i+K_{22}^i}{l_i^2}\sin\left(\mathbf{k}\cdot\mathbf{x}_i\right)\right\}\mathbf{t}_i\ , \tag{12}$$
$$C=\sum_{i=1}^{N}\left[K_{11}^i+K_{22}^i+2K_{12}^i\cos\left(\mathbf{k}\cdot\mathbf{x}_i\right)\right]\ .$$

When considering centrosymmetric cells, where to the *i*-th ligament with nodal stiffnesses $K_n^i$, $K_{11}^i$, $K_{12}^i$, $K_{22}^i$ corresponds the opposite *j*-th ligament with nodal stiffnesses $K_n^j=K_n^i$, $K_{11}^j=K_{22}^i=K_{22}^j=K_{11}^i$ and $K_{12}^j=K_{12}^i$, the terms in (12) take the simple form:

$$\mathbf{A} = 2\sum_{i=1}^{N}\left[K_n^i\left(\mathbf{n}_i \otimes \mathbf{n}_i\right) + 2\frac{K_{11}^i + K_{12}^i}{l_i^2}\left(\mathbf{t}_i \otimes \mathbf{t}_i\right)\right]\left[1 - \cos\left(\mathbf{k}\cdot\mathbf{x}_i\right)\right] ,$$

$$\mathbf{a} = \mathbf{0} ,$$

$$\mathbf{b} = 2\sum_{i=1}^{N}\left[\frac{K_{11}^i + K_{12}^i}{l_i}\sin\left(\mathbf{k}\cdot\mathbf{x}_i\right)\mathbf{t}_i\right] , \qquad (13)$$

$$C = 2\sum_{i=1}^{N}\left[K_{11}^i + K_{12}^i\cos\left(\mathbf{k}\cdot\mathbf{x}_i\right)\right] .$$

Finally, in case of ligaments with uniform section width *t*, one obtains:

$$\mathbf{A} = E_s\sum_{i=1}^{n}\left(\frac{t}{l_i}\right)\left[\left(\mathbf{n}_i \otimes \mathbf{n}_i\right) + \left(\frac{t}{l_i}\right)^2\left(\mathbf{t}_i \otimes \mathbf{t}_i\right)\right]\left[1 - \cos\left(\mathbf{k}\cdot\mathbf{x}_i\right)\right] ,$$

$$\mathbf{a} = \mathbf{0} ,$$

$$\mathbf{b} = \frac{1}{2}E_s\sum_{i=1}^{n}\left(\frac{t}{l_i}\right)^3 l_i\sin\left(\mathbf{k}\cdot\mathbf{x}_i\right)\mathbf{t}_i , \qquad (14)$$

$$C = \frac{1}{6}E_s\sum_{i=1}^{n}\left(\frac{t}{l_i}\right)^3 l_i^2\left[2 + \cos\left(\mathbf{k}\cdot\mathbf{x}_i\right)\right] .$$

The angular frequency $\omega(\mathbf{k})$, in the following called dispersive function, and the polarization vector $\hat{\mathbf{U}}(\mathbf{k})$ of a travelling wave with wave vector **k** are obtained by solving the eigenvalue problem (11), from which three dispersive branches are obtained. In the long wavelength limit $|\mathbf{k}| \to 0$, for which $\mathbf{A} = \mathbf{0}$, $\mathbf{a} = \mathbf{b} = \mathbf{0}$ and $C_0 = C\left(|\mathbf{k}| \to 0\right) = \sum_{i=1}^{N}\left[K_{11}^i + K_{22}^i + 2K_{12}^i\right]$, a double vanishing solution is obtained, from which two acoustic branches depart. A further non-vanishing solution $\omega_{opt} = \sqrt{\frac{C_0}{J}}$ is obtained that defines a critical point in the band structure, having vanishing group velocity ($v_g = \left.\frac{d\omega(\mathbf{k})}{d|\mathbf{k}|}\right|_{|\mathbf{k}|\to 0} = 0$), from which an optical branch departs.

## 3. Micropolar dynamic model for periodic lattices

The wave propagation in the discrete model analysed in the previous section may be approximate through the homogeneous continuum model here presented. As it will be shown in the examples of Section 4, the derived continuum is proved to be reliable also in case of intermediate wave-length in comparison to the characteristic length of the lattice.

### *3.1 Continualization method*

The discrete equation of motion (5) and (6) are transformed into the equation of motion of a continuum model by replacing the dof's of the Lagrangian model with a continuous generalized displacement field $\mathbf{v}(\mathbf{x},t)$ and $\theta(\mathbf{x},t)$. This field represents the generalized displacement of the reference cell located at $\mathbf{x}$ at time $t$ so that the displacement of the $i$-th adjacent cell is assumed according to the following second-order expansions

$$
\begin{aligned}
\mathbf{u}_i(t) &= \mathbf{v}(\mathbf{x},\mathrm{t}) + l_i\,\mathbf{H}(\mathbf{x},\mathrm{t})\mathbf{n}_i + \frac{1}{2} l_i^2 \nabla\mathbf{H}(\mathbf{x},\mathrm{t}) : (\mathbf{n}_i \otimes \mathbf{n}_i) + \mathcal{O}(l^3) \ , \\
\phi_i(t) &= \theta(\mathbf{x},\mathrm{t}) + l_i\,\boldsymbol{\chi}(\mathbf{x},\mathrm{t}) \cdot \mathbf{n}_i + \frac{1}{2} l_i^2 \nabla\boldsymbol{\chi}(\mathbf{x},\mathrm{t}) : (\mathbf{n}_i \otimes \mathbf{n}_i) + \mathcal{O}(l^3) \ ,
\end{aligned}
\tag{15}
$$

$\mathbf{H} = \nabla\mathbf{v}$ and $\nabla\mathbf{H}$ being the displacement gradient and the second gradient, $\boldsymbol{\chi} = \nabla\theta$ and $\nabla\boldsymbol{\chi}$ the curvature and its gradient tensor, respectively (see for instance Suiker *et al.*, 2001, Vasiliev *et al.*, 2010). An approximate formulation of the discrete equation of motion (5) and (6) is now obtained by considering the second order expansion of (15), namely

$$
\begin{aligned}
&\sum_{i=1}^{n} \left\{ \begin{array}{l} \left[ K_n^i (\mathbf{n}_i \otimes \mathbf{n}_i) + \dfrac{K_{11}^i + 2K_{12}^i + K_{22}^i}{l_i^2} (\mathbf{t}_i \otimes \mathbf{t}_i) \right] \left( l_i\,\mathbf{H}\mathbf{n}_i + \dfrac{1}{2} l_i^2 \nabla\mathbf{H} : (\mathbf{n}_i \otimes \mathbf{n}_i) \right) + \\ \qquad - \dfrac{K_{11}^i + K_{12}^i}{l_i} \mathbf{t}_i \theta - \dfrac{K_{12}^i + K_{22}^i}{l_i} \mathbf{t}_i \left\{ \theta + l_i\,\boldsymbol{\chi} \cdot \mathbf{n}_i + \dfrac{1}{2} l_i^2 \nabla\boldsymbol{\chi} : (\mathbf{n}_i \otimes \mathbf{n}_i) \right\} \end{array} \right\} - \mathrm{M}\ddot{\mathbf{v}} = \mathbf{0} \ , \\
&\sum_{i=1}^{n} \left[ \begin{array}{l} \dfrac{K_{11}^i + K_{12}^i}{l_i} \mathbf{t}_i \bullet \left( l_i\,\mathbf{H}\mathbf{n}_i + \dfrac{1}{2} l_i^2 \nabla\mathbf{H} : (\mathbf{n}_i \otimes \mathbf{n}_i) \right) + \\ \qquad - K_{11}^i \theta - K_{12}^i \left\{ \theta + l_i\,\boldsymbol{\chi} \cdot \mathbf{n}_i + \dfrac{1}{2} l_i^2 \nabla\boldsymbol{\chi} : (\mathbf{n}_i \otimes \mathbf{n}_i) \right\} \end{array} \right] - J\ddot{\theta} = 0 \ .
\end{aligned}
\tag{16}
$$

Since it has been assumed that the ligaments $i$ and $j = N + i$ have opposite direction, it follows $\mathbf{n}_j = -\mathbf{n}_i$ and $\mathbf{t}_j = -\mathbf{t}_i$, some terms in (16) turn out to be vanishing, namely

$$\sum_{i=1}^{n}\left\{K_n^i\left(\mathbf{n}_i\otimes\mathbf{n}_i\right)\mathbf{H}\mathbf{n}_i\right\}=\mathbf{0},\ \sum_{i=1}^{n}\left\{\frac{K_{11}^i+2K_{12}^i+K_{22}^i}{l_i}\left(\mathbf{t}_i\otimes\mathbf{t}_i\right)\mathbf{H}\mathbf{n}_i\right\}=\mathbf{0},\ \sum_{i=1}^{n}\left\{\frac{K_{11}^i+2K_{12}^i+K_{22}^i}{l_i}\mathbf{t}_i\theta\right\}=\mathbf{0}$$

and $\sum_{i=1}^{n}\left[K_{12}^i l_i\,\boldsymbol{\chi}\cdot\mathbf{n}_i\right]=0$ , and the equation of motion may be rewritten as follows

$$\begin{aligned}&\sum_{i}^{N}\left[K_n^i l_i^2\left(\mathbf{n}_i\otimes\mathbf{n}_i\otimes\mathbf{n}_i\otimes\mathbf{n}_i\right)+\left(K_{11}^i+2K_{12}^i+K_{22}^i\right)\left(\mathbf{t}_i\otimes\mathbf{t}_i\otimes\mathbf{n}_i\otimes\mathbf{n}_i\right)\right]\vdots\nabla\mathbf{H}+\\&\qquad-\sum_{i}^{N}\left(K_{11}^i+2K_{12}^i+K_{22}^i\right)\left(\mathbf{t}_i\otimes\mathbf{n}_i\right)\,\boldsymbol{\chi}+\sum_{i}^{N}\left[\frac{K_{11}^i-K_{22}^i}{2}l_i\,\left(\mathbf{t}_i\otimes\mathbf{n}_i\otimes\mathbf{n}_i\right)\right]:\nabla\boldsymbol{\chi}-\mathrm{M}\ddot{\mathbf{v}}=\mathbf{0}\ ,\end{aligned}\qquad(17)$$

$$\begin{aligned}&\sum_{i=1}^{N}\left[\left(K_{11}^i+2K_{12}^i+K_{22}^i\right)\left(\mathbf{t}_i\otimes\mathbf{n}_i\right)\right]:\mathbf{H}+\sum_{i=1}^{N}\left[\frac{K_{11}^i-K_{22}^i}{2}l_i\,\left(\mathbf{t}_i\otimes\mathbf{n}_i\otimes\mathbf{n}_i\right)\right]\vdots\nabla\mathbf{H}+\\&\qquad-\sum_{i=1}^{N}\left(K_{11}^i+2K_{12}^i+K_{22}^i\right)\,\theta-\sum_{i=1}^{N}\left[K_{12}^i l_i^2\left(\mathbf{n}_i\otimes\mathbf{n}_i\right)\right]:\nabla\boldsymbol{\chi}-J\ddot{\theta}=0\ .\end{aligned}\qquad(18)$$

In case of centrosymmetric lattice ( $K_{11}^i=K_{22}^i$ ), the above equations turn out to be similar to those obtained by Vasiliev *et al.*, 2010. Moreover, if the term $\mathcal{O}\left(l^2\right)$ in equation (18) are excluded, the rotation $\theta$ may be derived from (18) and after substituting in (17) one obtains a set of two equation of motion similar to those obtained by Gonella e Ruzzene, 2008.

By introducing the macro-rotation tensor $\mathbf{W}$ with components $w_{jh}=-\in_{3jh}\theta$, $\in_{jkl}$ being the Levi-Civita symbol, the following properties are obtained: $\theta=\mathbf{t}_i\cdot\mathbf{W}\mathbf{n}_i$, $\mathbf{n}_i\cdot\mathbf{W}\mathbf{n}_i=0$, $\left(\mathbf{t}_i\otimes\mathbf{t}_i\otimes\mathbf{n}_i\otimes\mathbf{n}_i\right)\vdots\nabla\mathbf{W}=\left(\mathbf{t}_i\otimes\mathbf{n}_i\right)\boldsymbol{\chi}$ and $\left(\mathbf{n}_i\otimes\mathbf{n}_i\otimes\mathbf{n}_i\otimes\mathbf{n}_i\right)\vdots\nabla\mathbf{W}=\mathbf{0}$. Therefore, once considered the Cosserat asymmetric strain tensor $\boldsymbol{\Gamma}=\mathbf{H}-\mathbf{W}\left(\theta\right)$, equation (17) may be rewritten in the form

$$\begin{aligned}&\sum_{i}^{N}\left[K_n^i l_i^2\left(\mathbf{n}_i\otimes\mathbf{n}_i\otimes\mathbf{n}_i\otimes\mathbf{n}_i\right)+\left(K_{11}^i+2K_{12}^i+K_{22}^i\right)\left(\mathbf{t}_i\otimes\mathbf{t}_i\otimes\mathbf{n}_i\otimes\mathbf{n}_i\right)\right]\vdots\nabla\boldsymbol{\Gamma}+\\&\qquad+\sum_{i}^{N}\left[\frac{K_{11}^i-K_{22}^i}{2}l_i\,\left(\mathbf{t}_i\otimes\mathbf{n}_i\otimes\mathbf{n}_i\right)\right]:\nabla\boldsymbol{\chi}-\mathrm{M}\ddot{\mathbf{v}}=\mathbf{0}\ .\end{aligned}\qquad(19)$$

Since the following relations apply $\left(\mathbf{t}_i\otimes\mathbf{t}_i\otimes\mathbf{n}_i\otimes\mathbf{n}_i\right)\vdots\nabla\boldsymbol{\Gamma}=\nabla\bullet\left[\left(\mathbf{t}_i\otimes\mathbf{n}_i\otimes\mathbf{t}_i\otimes\mathbf{n}_i\right):\boldsymbol{\Gamma}\right]$ and $\left(\mathbf{t}_i\otimes\mathbf{n}_i\otimes\mathbf{n}_i\right):\nabla\boldsymbol{\chi}=\nabla\bullet\left[\left(\mathbf{t}_i\otimes\mathbf{n}_i\otimes\mathbf{n}_i\right)\boldsymbol{\chi}\right]$, the equation of motion (19) may be rearranged in the

compact form

$$\nabla \bullet \left( \mathbb{E}_s \boldsymbol{\Gamma} + \mathbb{Y}_s \boldsymbol{\chi} \right) = \rho \ddot{\mathbf{v}} \ , \tag{20}$$

where the fourth and third order elasticity tensors of the equivalent homogeneous continuum are introduced, respectively, according to the following definition

$$\mathbb{E}_s = \frac{1}{A_{cell}} \sum_i^N \left[ K_n^i l_i^2 \left( \mathbf{n}_i \otimes \mathbf{n}_i \otimes \mathbf{n}_i \otimes \mathbf{n}_i \right) + \left( K_{11}^i + 2K_{12}^i + K_{22}^i \right) \left( \mathbf{t}_i \otimes \mathbf{n}_i \otimes \mathbf{t}_i \otimes \mathbf{n}_i \right) \right] \ , \tag{21}$$

$$\mathbb{Y}_s = \frac{1}{A_{cell}} \sum_i^N \left[ \frac{K_{11}^i - K_{22}^i}{2} l_i \ \left( \mathbf{t}_i \otimes \mathbf{n}_i \otimes \mathbf{n}_i \right) \right] \ , \tag{22}$$

$A_{cell}$ being the area of the periodic cell and $\rho = M/A_{cell}$ the mass density of the equivalent homogeneous continuum. It is worth to note that the form (21) guarantees that the fourth order tensor is endowed of the major symmetry. Moreover, from equation (20), the asymmetric macro-stress tensor

$$\mathbf{T} = \mathbb{E}_s \boldsymbol{\Gamma} + \mathbb{Y}_s \boldsymbol{\chi} \ , \tag{23}$$

may be identified, having components

$$\begin{aligned} \sigma_{hk} = & \frac{1}{A_{cell}} \sum_i^N \left[ K_n^i l_i^2 \left( n_h^i n_k^i n_p^i n_q^i \right) + \left( K_{11}^i + 2K_{12}^i + K_{22}^i \right) \left( t_h^i n_k^i t_p^i n_q^i \right) \right] \gamma_{pq} + \\ & + \frac{1}{A_{cell}} \sum_i^N \left[ \frac{K_{11}^i - K_{22}^i}{2} l_i \ \left( t_h^i n_k^i n_q^i \right) \right] \chi_q \quad . \end{aligned} \tag{24}$$

Let consider now the in-plane couple due to the shearing stresses

$$\begin{aligned} \sigma_{21} - \sigma_{12} = & - \in_{3jh} \left( \mathbf{e}_j \otimes \mathbf{e}_h \right) : \left( \mathbb{E}_s \boldsymbol{\Gamma} + \mathbb{Y}_s \boldsymbol{\chi} \right) = \\ & = \frac{1}{A_{cell}} \sum_i^N \left[ \left( K_{11}^i + 2K_{12}^i + K_{22}^i \right) t_p^i n_q^i \right] \gamma_{pq} + \frac{1}{A_{cell}} \sum_i^N \left[ \frac{K_{11}^i - K_{22}^i}{2} l_i \, n_q^i \right] \chi_q = \\ & = \frac{1}{A_{cell}} \sum_i^N \left( K_{11}^i + 2K_{12}^i + K_{22}^i \right) \left[ \left( \mathbf{t}_i \otimes \mathbf{n}_i \right) : \mathbf{H} - \theta \right] + \frac{1}{A_{cell}} \sum_i^N \left[ \frac{K_{11}^i - K_{22}^i}{2} l_i \, \mathbf{n}_i \right] \bullet \boldsymbol{\chi} \quad . \end{aligned} \tag{25}$$

By comparing this equation with the equation of motion (18) one obtains

$$
\frac{1}{A_{cell}}\sum_{i=1}^{N}\left\{\frac{K_{11}^{i}-K_{22}^{i}}{2}l_i\left[\left(\mathbf{t}_i\otimes\mathbf{n}_i\otimes\mathbf{n}_i\right)\vdots\nabla\mathbf{H}-\mathbf{n}_i\bullet\boldsymbol{\chi}\right]\right\}-\frac{1}{A_{cell}}\sum_{i=1}^{N}\left[K_{12}^{i}l_i^{2}\left(\mathbf{n}_i\otimes\mathbf{n}_i\right)\right]:\nabla\boldsymbol{\chi}+ \\
-\in_{3jh}\left(\mathbf{e}_j\otimes\mathbf{e}_h\right):\left(\mathbb{E}_s\boldsymbol{\Gamma}+\mathbb{Y}_s\boldsymbol{\chi}\right)-I\ddot{\theta}=0\ , \tag{26}
$$

where the density of rotational inertia $I = J/A_{cell}$ has been defined. From the following identity $\left(\mathbf{t}\otimes\mathbf{n}\otimes\mathbf{n}\right)\vdots\nabla\mathbf{H}-\mathbf{n}\bullet\boldsymbol{\chi}=\left(\mathbf{t}\otimes\mathbf{n}\otimes\mathbf{n}\right)\vdots\nabla\boldsymbol{\Gamma}=\nabla\bullet\left[\left(\mathbf{n}\otimes\mathbf{t}\otimes\mathbf{n}\right):\boldsymbol{\Gamma}\right]$ and noting from definition (22) that

$$
\mathbb{Y}_s^{T}=\frac{1}{A_{cell}}\sum_{i}^{N}\left[\frac{K_{11}^{i}-K_{22}^{i}}{2}l_i\left(\mathbf{n}_i\otimes\mathbf{t}_i\otimes\mathbf{n}_i\right)\right]\ , \tag{27}
$$

the first term in (26) may be identified as $\nabla\bullet\left(\mathbb{Y}_s^{T}\ \boldsymbol{\Gamma}\right)$. Moreover, once defined the second order elasticity tensor

$$
\mathbf{E}_s=-\frac{1}{A_{cell}}\sum_{i=1}^{N}\left[K_{12}^{i}l_i^{2}\left(\mathbf{n}_i\otimes\mathbf{n}_i\right)\right]\ , \tag{28}
$$

symmetric and negative defined, the second term in (26) may be identified as $\nabla\bullet\left(\mathbf{E}_s\ \boldsymbol{\chi}\right)$. Therefore, the second equation of motion (26) takes the form

$$
\nabla\bullet\left(\mathbb{Y}_s^{T}\ \boldsymbol{\Gamma}+\mathbf{E}_s\ \boldsymbol{\chi}\right)-\in_{3jh}\left(\mathbf{e}_j\otimes\mathbf{e}_h\right):\left(\mathbb{E}_s\boldsymbol{\Gamma}+\mathbb{Y}_s\boldsymbol{\chi}\right)=I\ddot{\theta}\quad,\quad j,h=1,2\quad, \tag{29}
$$

from which the second constitutive equation may be identified

$$
\mathbf{m}=\mathbb{Y}_s^{T}\ \boldsymbol{\Gamma}+\mathbf{E}_s\ \boldsymbol{\chi}\ , \tag{30}
$$

that provides the couple-stress vector of the equivalent continuum. The equation of motion (20), (29) and the constitutive equations (23) and (30) correspond to those ones of an equivalent micropolar model defined in terms of the generalized displacement field $\mathbf{v}(\mathbf{x},t)$ and $\theta(\mathbf{x},t)$, the asymmetric strain tensor $\boldsymbol{\Gamma}(\mathbf{x},t)$ and the curvature vector $\boldsymbol{\chi}(\mathbf{x},t)$, the asymmetric stress tensor $\mathbf{T}(\mathbf{x},t)$ and the couple-stress vector $\mathbf{m}(\mathbf{x},t)$. In general, a model without centro-symmetry is obtained in which the coupling between asymmetrical strains and curvatures is ruled by the elasticity tensor $\mathbb{Y}_s$. This tensor vanishes in case of centrosymmetric cells, a circumstance which is realized in the adopted cell topology when, for each pair of opposites ligaments, the stiffness terms are assumed to be equal $K_{11}^{i}=K_{22}^{i}=K_{11}^{j}=K_{22}^{j}$.

Finally, it must be emphasized that the second order elasticity tensor $\mathbf{E}_s$ coupling the couple-stresses to the curvatures in the micropolar continuum is negative defined as shown from equation (28). Bazant and Christensen (1972) highlighted this point in deriving a homogenized micropolar continuum for rectangular trusses and proposed an interpretation based on the generalized macro-homogeneity criterion, later applied by Kumar and McDowell (2004). In the next sub-section this approach is extended to non-centrosymmetric lattices.

*3.2 Extended Hamiltonian derivation of the micropolar model*

Let us consider a cluster *C* of periodic cells having characteristic size $L \gg l$ with respect to the characteristic cell size *l*. The associated Lagrangian function is $\mathcal{L}_C = T_C - \Pi_C$, where the kinetic energy is written in the form

$$T_C = \int_C \left[ \frac{1}{2}\rho \dot{\mathbf{v}}\cdot\dot{\mathbf{v}} + \frac{1}{2} I \dot{\theta}^2 \right] da \,, \tag{31}$$

being $\rho$ the average mass density and $I$ the density of rotational inertia already defined. The strain energy $\Pi_C$ stored in the cluster is obtained as the superposition of the elastic potential energy stored in the ligaments, which is due to the ligament extension and bending, respectively. The elastic potential energy stored in the ligament connecting the reference cell centered in $\mathbf{x}$ to the *i*-th adjacent one is approximated through the second order expansion of the generalized displacement field (15). The axial extension of the *i*-th ligament is approximated as

$$\Delta_{di} \simeq l_i \mathbf{H} : \left(\mathbf{n}_i \otimes \mathbf{n}_i\right) + \frac{l_i^2}{2} \nabla\mathbf{H} \vdots \left(\mathbf{n}_i \otimes \mathbf{n}_i \otimes \mathbf{n}_i\right) = l_i \left(\mathbf{n}_i \otimes \mathbf{n}_i\right) : \mathbf{\Gamma} + \frac{l_i^2}{2} \left(\mathbf{n}_i \otimes \mathbf{n}_i \otimes \mathbf{n}_i\right) \vdots \nabla\mathbf{\Gamma} \,, \tag{32}$$

being $\mathbf{n}_i \cdot \mathbf{W}\mathbf{n}_i = 0$ and the symbol $\vdots$ denoting the triple scalar product (i.e. $\mathbf{A} \vdots \left(\mathbf{b} \otimes \mathbf{c} \otimes \mathbf{d}\right) = a_{i,jh} b_i c_j d_h$ ). Moreover, the transverse displacement is approximated as

$$\Delta_{ti} \simeq l_i \mathbf{H} : \left(\mathbf{t}_i \otimes \mathbf{n}_i\right) + \frac{l_i^2}{2} \nabla\mathbf{H} \vdots \left(\mathbf{t}_i \otimes \mathbf{n}_i \otimes \mathbf{n}_i\right) . \tag{33}$$

The approximations of the end rotations are derived by noting that $\phi = \mathbf{t}_i \cdot \mathbf{W}\mathbf{n}_i$ and $\nabla\mathbf{W}\vdots(\mathbf{t}_i\otimes\mathbf{n}_i\otimes\mathbf{n}_i)=\mathbf{n}\bullet\boldsymbol{\chi}$, and depend on the tensor of asymmetric strain, on the curvature and on their gradients in the forms

$$\begin{aligned}\varphi=\phi-\frac{\Delta_{ti}}{l_i}&=-(\mathbf{t}_i\otimes\mathbf{n}_i):\boldsymbol{\Gamma}-\frac{l_i}{2}\nabla\mathbf{H}\vdots(\mathbf{t}_i\otimes\mathbf{n}_i\otimes\mathbf{n}_i),\\ \varphi_i=\phi_i-\frac{\Delta_{ti}}{l_i}&=-(\mathbf{t}_i\otimes\mathbf{n}_i):\boldsymbol{\Gamma}-\frac{l_i}{2}\nabla\mathbf{H}\vdots(\mathbf{t}_i\otimes\mathbf{n}_i\otimes\mathbf{n}_i)+l_i\boldsymbol{\chi}\cdot\mathbf{n}_i+\frac{l_i^2}{2}\nabla\boldsymbol{\chi}:(\mathbf{n}_i\otimes\mathbf{n}_i).\end{aligned}\tag{34}$$

The elastic potential energy stored in the cluster due to the ligaments extension in the cells is derived by equation (2) and takes the form

$$\Pi_{Ca}\simeq\frac{1}{2}\int_C\sum_{i=1}^{n}\frac{1}{2A_{cell}}K_n^i l_i^2\left[(\mathbf{n}_i\otimes\mathbf{n}_i):\boldsymbol{\Gamma}+\frac{l_i}{2}(\mathbf{n}_i\otimes\mathbf{n}_i\otimes\mathbf{n}_i)\vdots\nabla\boldsymbol{\Gamma}\right]^2 da\ .\tag{35}$$

By disregarding the term depending on the gradient of the tensor of asymmetric strain, this contribution to the elastic potential energy takes the following quadratic form in $\boldsymbol{\Gamma}$

$$\Pi_{Ca}\simeq\frac{1}{2}\int_C\boldsymbol{\Gamma}:\left[\sum_{i=1}^{n}\frac{1}{2A_{cell}}K_n^i l_i^2(\mathbf{n}_i\otimes\mathbf{n}_i\otimes\mathbf{n}_i\otimes\mathbf{n}_i)\right]\boldsymbol{\Gamma}\ da\ .\tag{36}$$

Likewise, by substituting (34) in (3), one obtains the elastic potential energy stored in the cluster depending on the bending of the ligaments connecting the reference cell to the *i*-th surrounding one

$$\Pi_{Cb}\cong\frac{1}{2}\int_C\sum_{i=1}^{n}\frac{1}{2A_{cell}}\left\{\begin{aligned}&K_{11}^i\left[-(\mathbf{t}_i\otimes\mathbf{n}_i):\boldsymbol{\Gamma}-\frac{l_i}{2}\nabla\mathbf{H}\vdots(\mathbf{t}_i\otimes\mathbf{n}_i\otimes\mathbf{n}_i)\right]^2+\\&+2K_{12}^i\left[-(\mathbf{t}_i\otimes\mathbf{n}_i):\boldsymbol{\Gamma}-\frac{l_i}{2}\nabla\mathbf{H}\vdots(\mathbf{t}_i\otimes\mathbf{n}_i\otimes\mathbf{n}_i)\right]\cdot\\&\left[-(\mathbf{t}_i\otimes\mathbf{n}_i):\boldsymbol{\Gamma}-\frac{l_i}{2}\nabla\mathbf{H}\vdots(\mathbf{t}_i\otimes\mathbf{n}_i\otimes\mathbf{n}_i)+l_i\boldsymbol{\chi}\cdot\mathbf{n}_i+\frac{l_i^2}{2}\nabla\boldsymbol{\chi}:(\mathbf{n}_i\otimes\mathbf{n}_i)\right]+\\&K_{22}^i\left[-(\mathbf{t}_i\otimes\mathbf{n}_i):\boldsymbol{\Gamma}-\frac{l_i}{2}\nabla\mathbf{H}\vdots(\mathbf{t}_i\otimes\mathbf{n}_i\otimes\mathbf{n}_i)+l_i\boldsymbol{\chi}\cdot\mathbf{n}_i+\frac{l_i^2}{2}\nabla\boldsymbol{\chi}:(\mathbf{n}_i\otimes\mathbf{n}_i)\right]^2\end{aligned}\right\}da\ .\tag{37}$$

Because the assumption on the stiffnesses of opposite ligaments *i*-th and *j*-th ( $j=N+i$ ) with $\mathbf{x}_j=-\mathbf{x}_i$ and $\mathbf{t}_j=-\mathbf{t}_i$, namely $K_n^j=K_n^i$, $K_{11}^j=K_{22}^i$, $K_{22}^j=K_{11}^i$, $K_{12}^j=K_{12}^i$, the summations over *n* of dyadic products of an odd number of vectors $\mathbf{n}_i$, $\mathbf{t}_i$ in (37) are equal to zero. Since the continuum model is here limited to the classical micropolar form (see for reference Altenbach *et al.*, 2011), i.e. to a first gradient model, the terms in (35) and (37) depending on the second gradients of the generalized displacement $\nabla \mathbf{H}$ and $\nabla \boldsymbol{\chi}$ are disregarded, apart from the term in (37) involving the product of $\mathbf{W}$ ($\boldsymbol{\Gamma} = \mathbf{H} - \mathbf{W}$) and $\nabla \boldsymbol{\chi}$, namely $\left[(\mathbf{t}_i \otimes \mathbf{n}_i) : \mathbf{W}\right]\left[\frac{l_i^2}{2} \nabla \boldsymbol{\chi} : (\mathbf{n}_i \otimes \mathbf{n}_i)\right]$. Then, the total potential energy in the cluster due to the ligament bending may be written as follows

$$\Pi_{Cb} \simeq \frac{1}{2}\int_C \sum_{i=1}^{N} \frac{1}{2A_{cell}} \left\{ \begin{array}{l} 2\left(K_{11}^i + 2K_{12}^i + K_{22}^i\right)\left[(\mathbf{t}_i \otimes \mathbf{n}_i) : \boldsymbol{\Gamma}\right]^2 + \\ +2K_{12}^i l_i^2 \left[(\mathbf{t}_i \otimes \mathbf{n}_i) : \mathbf{W}\right] \cdot \left[\nabla \boldsymbol{\chi} : (\mathbf{n}_i \otimes \mathbf{n}_i)\right] + \\ +\left(K_{11}^i + K_{22}^i\right) l_i^2 \left[\boldsymbol{\chi} \cdot \mathbf{n}_i\right]^2 + \\ +\left(K_{11}^i + K_{22}^i\right) l_i^2 \left[(\mathbf{t}_i \otimes \mathbf{n}_i) : \mathbf{W}\right]\left[\nabla \boldsymbol{\chi} : (\mathbf{n}_i \otimes \mathbf{n}_i)\right] + \\ +2\left(K_{11}^i - K_{22}^i\right) l_i \, \boldsymbol{\Gamma} : (\mathbf{t}_i \otimes \mathbf{n}_i \otimes \mathbf{n}_i)\boldsymbol{\chi} \end{array} \right\} da \ . \tag{38}$$

Noting that the second and fourth terms in (38) may be written as $\left[(\mathbf{t}_i \otimes \mathbf{n}_i) : \mathbf{W}\right]\left[\nabla \boldsymbol{\chi} : (\mathbf{n}_i \otimes \mathbf{n}_i)\right] = \theta \ \nabla \boldsymbol{\chi} : (\mathbf{n}_i \otimes \mathbf{n}_i) = \theta \ \theta_{,pq} n_p^i n_q^i$, by the application of the Divergence theorem (see Bažant and Christensen (1972) and Kumar and McDowell (2004)) one obtains

$$\int_C \sum_{i=1}^{N} \left[(\mathbf{t}_i \otimes \mathbf{n}_i) : \mathbf{W}\right]\left[\nabla \boldsymbol{\chi} : (\mathbf{n}_i \otimes \mathbf{n}_i)\right] da \ = \sum_{i}^{N} n_p^i n_q^i \int_C \theta \ \theta_{,pq} \ da =$$

$$= -\sum_{i=1}^{N} n_p^i n_q^i \int_C \theta_{,p} \ \theta_{,q} \ da + \sum_{i=1}^{N} n_p^i n_q^i \int_{\partial C} \theta_{,p} \ \theta \ \nu_q \ \ ds = -\sum_{i=1}^{N} \int_C (\boldsymbol{\chi} \cdot \mathbf{n}_i)^2 \ da + \text{boundary terms} \ ,$$

being the boundary terms defined on the boundary of *C*. The bending elastic potential energy in the cluster is written in the quadratic form involving the asymmetric strain and the curvature

$$\Pi_{Cb} = \frac{1}{2}\int_C \sum_{i=1}^{N} \frac{1}{A_{cell}} \left\{ \begin{array}{l} \left(K_{11}^i + 2K_{12}^i + K_{22}^i\right)\left[(\mathbf{t}_i \otimes \mathbf{n}_i) : \boldsymbol{\Gamma}\right]^2 - K_{12}^i l_i^2 \left[\boldsymbol{\chi} \cdot \mathbf{n}_i\right]^2 + \\ +\left(K_{11}^i - K_{22}^i\right) l_i \, \boldsymbol{\Gamma} : (\mathbf{t}_i \otimes \mathbf{n}_i \otimes \mathbf{n}_i)\boldsymbol{\chi} \end{array} \right\} da \ . \tag{39}$$

By summing the elastic potential energy due to the axial extension (36) to that associated to the bending (39) of the ligaments, the elastic potential energy of the cluster C is obtained, which results in the classical form of the non-centrosymmetric micropolar continuum

$$\Pi_C = \frac{1}{2}\int_C \left( \boldsymbol{\Gamma} : \mathbb{E}_s \boldsymbol{\Gamma} + \boldsymbol{\chi} \cdot \mathbf{E}_s \boldsymbol{\chi} + 2\, \boldsymbol{\Gamma} : \mathbb{Y}_s \boldsymbol{\chi} \right) da \quad , \tag{40}$$

with the elasticity tensors given by (21), (22) and (28).

It must be remarked that assuming a first order expansion of the generalized displacements in (15) implies the same fourth (21) and third order tensors (22), while the second order tensor, associated to the curvature vector, differs from (28) and results to be positive definite in the form

$$\mathbf{E}_s^{+} = \frac{1}{2A_{cell}} \sum_{i=1}^{N} \left[ \left( K_{11}^{i} + K_{22}^{i} \right) l_i^2 \left( \mathbf{n}_i \otimes \mathbf{n}_i \right) \right] . \tag{41}$$

The Lagrangian takes the following form

$$\mathcal{L} = \frac{1}{2}\int_C \left[ \rho \dot{\mathbf{v}} \cdot \dot{\mathbf{v}} + I \dot{\theta}^2 - \boldsymbol{\Gamma} \bullet \mathbb{E}_s \boldsymbol{\Gamma} - \boldsymbol{\chi} \cdot \mathbf{E}_s \boldsymbol{\chi} - 2\, \boldsymbol{\Gamma} : \mathbb{Y}_s \boldsymbol{\chi} \right] da + \text{boundary terms} \; , \tag{42}$$

where $C$ is the considered cluster of cells. By applying the extended Hamilton principle, the equation of motion of the micropolar continuum is obtained independently on the boundary terms in equation (42) and are those of the micropolar continuum already obtained in equations (20) and (29).

From the definition (40), the density $\pi_s$ of the elastic potential energy is obtained, and hence the constitutive equations

$$\mathbf{T} = \frac{\partial \pi_s}{\partial \boldsymbol{\Gamma}} = \mathbb{E}_s \boldsymbol{\Gamma} + \mathbb{Y}_s \boldsymbol{\chi} \quad , \qquad \mathbf{m} = \frac{\partial \pi_s}{\partial \boldsymbol{\chi}} = \mathbf{E}_s \boldsymbol{\chi} + \mathbb{Y}_s^T \boldsymbol{\Gamma} \; . \tag{43}$$

The components of the asymmetric stress tensor $\mathbf{T}$ are denoted with $\sigma_{11}, \sigma_{12}, \sigma_{21}, \sigma_{22}$, while those of the vector of couple stress $\mathbf{m}$ are denoted with $m_1$ e $m_2$, which are energetically conjugated to the components $\gamma_{11} = u_{1,1}$, $\gamma_{22} = u_{2,2}$, $\gamma_{12} = u_{1,2} + \phi$, $\gamma_{21} = u_{2,1} - \phi$ of the asymmetric strain tensor $\boldsymbol{\Gamma}$ and to the components $\chi_1 = \phi_{,1}$ and $\chi_2 = \phi_{,2}$ of the curvature $\boldsymbol{\chi}$.

## 4. Harmonic wave propagation in the equivalent micropolar model

The propagation of elastic waves in the two-dimensional micropolar continuum is obtained considering the harmonic motion $\mathbf{v}=\hat{\mathbf{v}}\exp\left[i\left(\mathbf{k}\cdot\mathbf{x}-\omega t\right)\right]$ and $\theta=\hat{\theta}\exp\left[i\left(\mathbf{k}\cdot\mathbf{x}-\omega t\right)\right]$, where the polarization vector is defined by the collection of $\hat{\mathbf{v}}$ and $\hat{\theta}$. From this assumption it follows $\mathbf{H}=\hat{\mathbf{v}}\otimes\mathbf{k}\exp\left[i\left(\mathbf{k}\cdot\mathbf{x}-\omega t\right)\right]$, $\nabla\mathbf{H}=-\hat{\mathbf{v}}\otimes\mathbf{k}\otimes\mathbf{k}\exp\left[i\left(\mathbf{k}\cdot\mathbf{x}-\omega t\right)\right]$ and $\boldsymbol{\chi}=i\hat{\theta}\mathbf{k}\exp\left[i\left(\mathbf{k}\cdot\mathbf{x}-\omega t\right)\right]$, $\nabla\boldsymbol{\chi}=i\hat{\theta}\mathbf{k}\otimes\mathbf{k}\exp\left[i\left(\mathbf{k}\cdot\mathbf{x}-\omega t\right)\right]$. The equation of motion (17) and (18) take the homogeneous linear form

$$
\begin{aligned}
&\frac{1}{A_{cell}}\sum_{i}^{N}\left[\left(\mathbf{n}_i\otimes\mathbf{n}_i\right):\left(\mathbf{k}\otimes\mathbf{k}\right)\right]\left[K_n^i l_i^2\left(\mathbf{n}_i\otimes\mathbf{n}_i\right)+\left(K_{11}^i+2K_{12}^i+K_{22}^i\right)\left(\mathbf{t}_i\otimes\mathbf{t}_i\right)\right]\hat{\mathbf{v}}+\\
&+\frac{1}{A_{cell}}\sum_{i}^{N}\left[\frac{K_{11}^i-K_{22}^i}{2}l_i\left[\left(\mathbf{n}_i\otimes\mathbf{n}_i\right):\left(\mathbf{k}\otimes\mathbf{k}\right)\right]\right]\mathbf{t}_i\hat{\theta}+i\frac{1}{A_{cell}}\sum_{i}^{N}\left(K_{11}^i+2K_{12}^i+K_{22}^i\right)\left(\mathbf{k}\bullet\mathbf{n}_i\right)\mathbf{t}_i\,\hat{\theta}-\omega^2\rho\hat{\mathbf{v}}=\mathbf{0}\ ,
\end{aligned}
\tag{44}
$$

$$
\begin{aligned}
&\frac{1}{A_{cell}}\sum_{i=1}^{N}\left\{\left[\frac{K_{11}^i-K_{22}^i}{2}l_i\left[\left(\mathbf{n}_i\otimes\mathbf{n}_i\right):\left(\mathbf{k}\otimes\mathbf{k}\right)\right]\right]\mathbf{t}_i\right\}\bullet\hat{\mathbf{v}}-i\frac{1}{A_{cell}}\sum_{i=1}^{N}\left\{\left[\left(K_{11}^i+2K_{12}^i+K_{22}^i\right)\left(\mathbf{k}\bullet\mathbf{n}_i\right)\right]\mathbf{t}_i\right\}\bullet\hat{\mathbf{v}}+\\
&\qquad+\frac{1}{A_{cell}}\sum_{i=1}^{N}\left\{\left(K_{11}^i+2K_{12}^i+K_{22}^i\right)-K_{12}^i l_i^2\left(\mathbf{n}_i\otimes\mathbf{n}_i\right):\left(\mathbf{k}\otimes\mathbf{k}\right)\right\}\hat{\theta}-\omega^2 I\hat{\theta}=0\ .
\end{aligned}
\tag{45}
$$

The eigenvalue problem $\mathbf{C}_{Hom}\left(\mathbf{k},\omega\right)\hat{\mathbf{V}}=\mathbf{0}$ is obtained, being $\hat{\mathbf{V}}=\left\{\hat{\mathbf{v}}^T\quad\hat{\theta}\right\}$, with the same structure of problem (11) and ruled by a hermitian matrix $\mathbf{C}_{Hom}\left(\mathbf{k},\omega\right)$ with submatrices

$$
\begin{aligned}
\mathbf{A}_{\mathrm{hom}}&=\frac{1}{A_{cell}}\sum_{i}^{N}\left[\left(\mathbf{n}_i\otimes\mathbf{n}_i\right):\left(\mathbf{k}\otimes\mathbf{k}\right)\right]\left[K_n^i l_i^2\left(\mathbf{n}_i\otimes\mathbf{n}_i\right)+\left(K_{11}^i+2K_{12}^i+K_{22}^i\right)\left(\mathbf{t}_i\otimes\mathbf{t}_i\right)\right]\ ,\\
\mathbf{a}_{\mathrm{hom}}&=\frac{1}{A_{cell}}\sum_{i}^{N}\left[\frac{K_{11}^i-K_{22}^i}{2}l_i\left[\left(\mathbf{n}_i\otimes\mathbf{n}_i\right):\left(\mathbf{k}\otimes\mathbf{k}\right)\right]\right]\mathbf{t}_i\ ,\\
\mathbf{b}_{\mathrm{hom}}&=\frac{1}{A_{cell}}\sum_{i}^{N}\left(K_{11}^i+2K_{12}^i+K_{22}^i\right)\left(\mathbf{k}\bullet\mathbf{n}_i\right)\mathbf{t}_i\ ,\\
C_{\mathrm{hom}}&=\frac{1}{A_{cell}}\sum_{i=1}^{N}\left[\left(K_{11}^i+2K_{12}^i+K_{22}^i\right)-K_{12}^i l_i^2\left(\mathbf{n}_i\otimes\mathbf{n}_i\right):\left(\mathbf{k}\otimes\mathbf{k}\right)\right]\ .
\end{aligned}
\tag{46}
$$

The secular equation provides three dispersion functions $\omega_h\left(\mathbf{k}\right)$, $h$=1,3. In the long-wave asymptotic, namely for $\left|\mathbf{k}\right|\to 0$, two acoustic branches are obtained together with an optical branch departing from a critical point at frequency $\omega_{opt}\left(|\mathbf{k}|\to 0\right)=\sqrt{C_{0,\hom}/I}=\sqrt{C_0/J}$ that equals the corresponding one from the discrete model described in Section 2. By comparing the submatrices (46) with those (12) and noting that $\cos\left(\mathbf{k}\cdot\mathbf{x}_i\right)\simeq 1-\frac{1}{2}\left(\mathbf{x}_i\otimes\mathbf{x}_i\right):\left(\mathbf{k}\otimes\mathbf{k}\right)$ and $\sin\left(\mathbf{k}\cdot\mathbf{x}_i\right)\simeq\mathbf{k}\cdot\mathbf{x}_i$, it follows that $\mathbf{C}_{Lag}\left(\mathbf{k},\omega\right)=A_{cell}\ \mathbf{C}_{\hom}\left(\mathbf{k},\omega\right)+\mathcal{O}\left(|\mathbf{k}|^3\right)$, therefore the hermitian matrix of the Lagrangian system $\mathbf{C}_{Lag}\left(\mathbf{k},\omega\right)$ may be approximated by the corresponding one from the micropolar homogenized model considering the second order expansion of the matrix in the wave vector **k**. It may be easily verified that if the elastic positive defined second order tensor $\mathbf{E}_s^+$ given by (41) is assumed, that is derived by a first order expansion of the rotation field, the optical branch turns out to be approximated by the equivalent micropolar continuum with a lower accuracy. This circumstance is shown in the examples of Section 5 where the possibility of loss of strong hyperbolicity in cases of wave propagation for the examples considered is investigated. The problem associated to the negative definiteness of the second order tensor remains open for the equilibrium problems of micropolar continua equivalent to the lattice, as remarked by Kumar and McDowell (2004) and taken up by Liu et al. (2012) for chiral lattices.

## 5. Examples

The sensitivity of the acoustic band structure and the formation of stop bands on the non-centrosymmetric topology of the lattice is analysed and discussed with reference to some examples regarding both square and triangular lattice. Based on the non-centrosymmetric lattice structure assumed in Section 2, here the ligaments are assumed having a one step change in the thickness at midspan as shown in Figures 4 and 12. Therefore, the thicknesses are denoted by $t$ and $\alpha t$, respectively. The axial stiffness of the ligaments is $K_n=\frac{2E\alpha}{\left(1+\alpha\right)}\frac{t}{l}$, while the bending stiffness of the $i$-th ligament are given in the following forms:

a) $K_{hk}^i=F_{hk}\left(\alpha\right)\frac{Et^3}{l}$ , with $F_{11}\left(\alpha\right)=\frac{2\left(7+\alpha^3\right)}{3\left(1+14\alpha^3+\alpha^6\right)}$ , $F_{12}\left(\alpha\right)=\frac{4\alpha^3\left(1+\alpha^3\right)}{3\left(1+14\alpha^3+\alpha^6\right)}$, $F_{22}\left(\alpha\right)=\frac{2\alpha^3\left(1+7\alpha^3\right)}{3\left(1+14\alpha^3+\alpha^6\right)}$ for ligaments $i=1,2$ of the square lattice (Figure 4.b) and for ligaments $i = 2, 4, 6$ of the equilateral triangular lattice (Figure 12.b), respectively;

b) $K_{hk}^{i} = F_{hk}\left(\frac{1}{\alpha}\right)\frac{Et^{3}}{l}$ for the ligaments $i = 3, 4$ of the square lattice and for ligaments $i = 1, 3, 5$ of the equilateral triangular lattice. In the examples the centrosymmetric case $(\alpha = 1)$ and two non-centrosymetric ones $(\alpha = 2, 4)$ are considered.

The acoustic band structures given by the discrete model formulated in Section 2 is represented in the Brillouin zone (see Brillouin, 1953), along the closed polygonal curve $\Upsilon$ with vertices identified by the values $\Xi_0$, $\Xi_1$, $\Xi_2$, of the arc-length $\Xi$ in the dimensionless plane $(k_1 l, k_2 l)$ (see Figures 5.c and 13.c). A compact spectral description is given in terms of the dimensionless frequency $\omega \big/ \sqrt{E_s d / M}$, the ratios $\frac{t}{l}$ and $J \big/ Ml^2 = \left( r \big/ l \right)^2$ .

Centrosymmetric topologies are firstly analyzed ($\alpha = 1$) to obtain reference results. Then non-centrosymmetric topologies have been analyzed for the thickness ratios $t/l$ and $\alpha$. According to the discrete model described in Section 2, two acoustic branches are obtained together with an optical one characterized by a critical point for long-wave limit $|\mathbf{k}| \to 0$ whose frequency is $\omega_{opt}\left(|\mathbf{k}| \to 0\right) = \sqrt{\sum_{i=1}^{N}\left[K_{11}^{i} + K_{22}^{i} + 2K_{12}^{i}\right] \Big/ J}$ . This angular frequency is increasing with the ligament thickness ratio $t/l$ and with the stiffness step ratio $\alpha$. Finally, some examples are considered to investigate the accuracy of the results provided by the micropolar model. As this model is inherently formulated for the case of moderately long waves, the comparison between the results by the discrete model and those by the equivalent micropolar one are shown in a homothetic sub-region of the reduced Brillouin zone (see Figure 5(d) and 13(d)).

### *5.1. Square lattice*

The square lattice and the periodic cell are shown in Figure 4. The geometrical data are $A_{cell} = l^2$, $\mathbf{n}_1 = -\mathbf{t}_2 = \mathbf{e}_1$ , $\mathbf{n}_2 = \mathbf{t}_1 = \mathbf{e}_2$, $\mathbf{t}_2 = -\mathbf{t}_4 = -\mathbf{e}_1$. Moreover, the bending stiffnesses for ligaments $i = 1, N$ are equal, namely $K_{hk}^i = K_{hk}$.

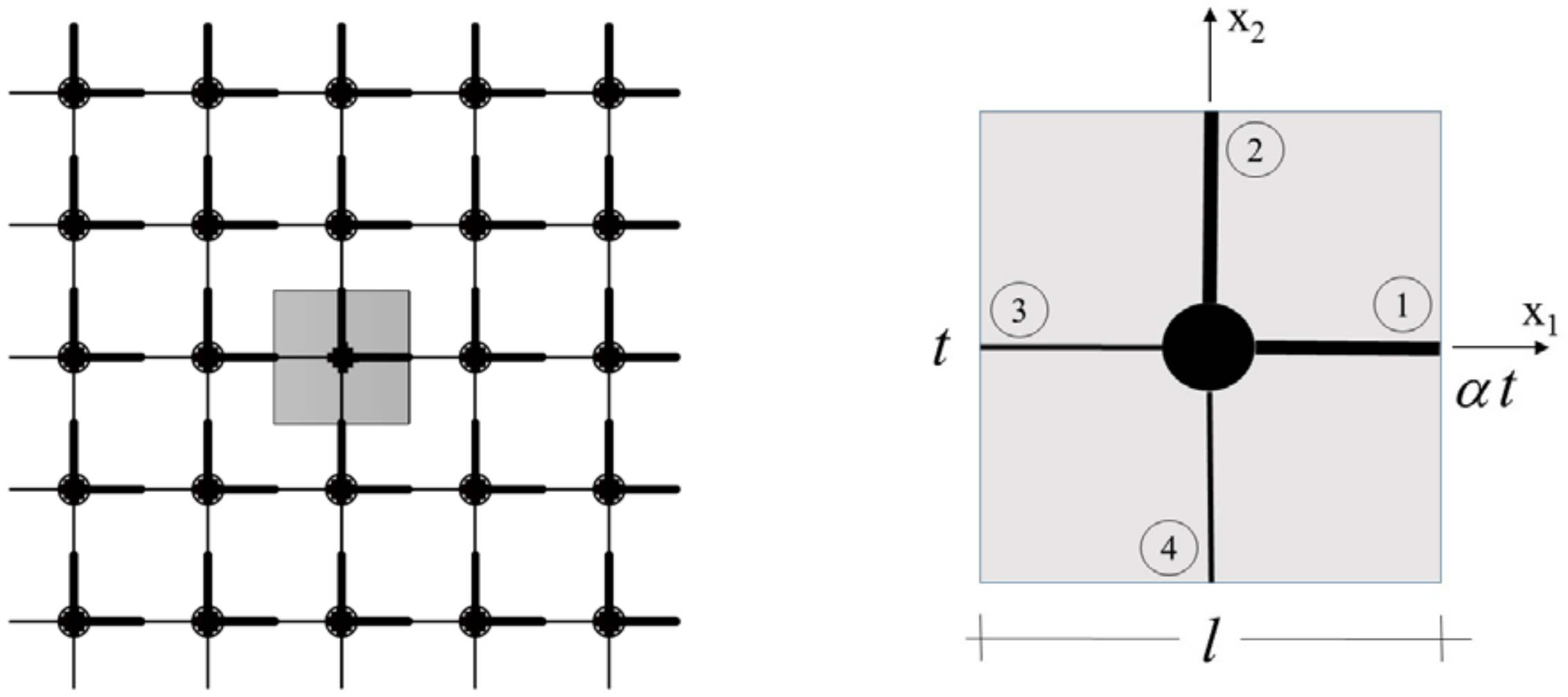

(a) (b)

Figure 4: Periodic non-centrosymmetric square lattice.

The harmonic wave propagation of the lagrangian model is obtained by solving the eigenvalue problem:

$$\begin{bmatrix} 2K_n\left[1-\cos\left(k_1 l\right)\right] + 2\dfrac{\hat{K}}{l^2}\left[1-\cos\left(k_2 l\right)\right] - \omega^2 M & 0 & -\dfrac{\tilde{K}}{l}\left[1-\cos\left(k_2 l\right)\right] - i\dfrac{\hat{K}}{l}\sin\left(k_2 l\right) \\ 0 & 2K_n\left[1-\cos\left(k_2 l\right)\right] + 2\dfrac{\hat{K}}{l^2}\left[1-\cos\left(k_1 l\right)\right] - \omega^2 M & \dfrac{\tilde{K}}{l}\left[1-\cos\left(k_1 l\right)\right] + i\dfrac{\hat{K}}{l}\sin\left(k_1 l\right) \\ -\dfrac{\tilde{K}}{l}\left[1-\cos\left(k_2 l\right)\right] + i\dfrac{\hat{K}}{l}\sin\left(k_2 l\right) & \dfrac{\tilde{K}}{l}\left[1-\cos\left(k_1 l\right)\right] - i\dfrac{\hat{K}}{l}\sin\left(k_1 l\right) & 2\left\{\hat{K} + K_{12}\left[\cos\left(k_1 l\right) + \cos\left(k_2 l\right) - 2\right]\right\} - \omega^2 J \end{bmatrix} \begin{Bmatrix} \hat{u}_1 \\ \hat{u}_2 \\ \hat{\phi} \end{Bmatrix} = \mathbf{0}\,, \tag{47}$$

where the parameters $\hat{K} = K_{11}^i + 2K_{12}^i + K_{22}^i$ and $\tilde{K} = \left(K_{11}^i - K_{22}^i\right)/2$ are introduced, so obtaining the dispersion function $\omega(\mathbf{k})$. In the long wave approximation, the constitutive equation of the micropolar continuum is obtained from (21), (22) and (28) and is written in the following matrix form (here the Voigt notation is assumed):

$$
\begin{Bmatrix} \sigma_{11} \\ \sigma_{22} \\ \sigma_{12} \\ \sigma_{21} \\ m_1 \\ m_2 \end{Bmatrix} = \begin{bmatrix} 2\mu & 0 & 0 & 0 & 0 & 0 \\ 0 & 2\mu & 0 & 0 & 0 & 0 \\ 0 & 0 & \kappa & 0 & 0 & -Y \\ 0 & 0 & 0 & \kappa & Y & 0 \\ 0 & 0 & 0 & Y & S & 0 \\ 0 & 0 & -Y & 0 & 0 & S \end{bmatrix} \begin{Bmatrix} \gamma_{11} \\ \gamma_{22} \\ \gamma_{12} \\ \gamma_{21} \\ \chi_1 \\ \chi_2 \end{Bmatrix} \quad , \tag{48}
$$

involving four elastic moduli $\mu = \frac{K_n}{2}$ , $\kappa = \frac{\hat{K}}{l^2}$ , $\mathrm{Y} = \frac{\tilde{K}}{l}$ , $S = -K_{12}$. If only the first order expansion of the generalized displacement field were considered in evaluating the total potential energy stored in the lattice, equation (41) applies and the elastic modulus involving the curvature takes the form $S^+ = \frac{(K_{11} + K_{22})}{2}$. In case of symmetric macro-strain fields $\gamma_{12} = \gamma_{21}$ with vanishing curvature $\boldsymbol{\chi} = \mathbf{0}$, the fourth order elastic tensor for the square lattice has the elastic moduli $C_{1111} = 2\mu$, $C_{1122} = 0$ and $C_{1212} = \kappa$. The approximate equivalent continuum formulation provides the eigenvalue problem written as follows:

$$
\begin{bmatrix} 2\mu k_1^2 + \kappa k_2^2 - \omega^2 \rho & 0 & -Yk_2^2 - i\kappa k_2 \\ 0 & 2\mu k_2^2 + \kappa k_1^2 - \omega^2 \rho & Yk_1^2 + i\kappa k_1 \\ -Yk_2^2 + i\kappa k_2 & Yk_1^2 - i\kappa k_1 & 2\left[\kappa + S\left(k_1^2 + k_1^2\right)\right] - \omega^2 I \end{bmatrix} \begin{Bmatrix} \hat{u}_1 \\ \hat{u}_2 \\ \hat{\phi} \end{Bmatrix} = \mathbf{0} \quad . \tag{49}
$$

In the long-wave limit $\lambda \to \infty$ ($|\mathbf{k}| \to 0$) the angular frequencies are $\omega_{aco1,2} = 0$ and $\omega_{opt} = \sqrt{\frac{2\kappa}{I}}$, the third one being depending on the non-centrosymmetry of the periodic cell being $\kappa = \frac{K_{11}^i + 2K_{12}^i + K_{22}^i}{l^2}$. Finally, in case of centrosymmetric cell, i.e. $\alpha = 1$ , the following elastic moduli are obtained $\mu = \frac{Et}{2l}$ , $\kappa = E\left(\frac{t}{l}\right)^3$, $Y = 0$, $S = -\frac{El^2}{6}\left(\frac{t}{l}\right)^3$ and the eigenproblem takes the form

$$\begin{bmatrix} k_1^2 + \varsigma k_2^2 - \omega^2 \dfrac{\rho l}{Et} & 0 & -i\varsigma k_2 \\ 0 & k_2^2 + \varsigma k_1^2 - \omega^2 \dfrac{\rho l}{Et} & i\varsigma k_1 \\ i\varsigma k_2 & -i\varsigma k_1 & 2\varsigma\left[1 - \dfrac{l^2}{6}\left(k_1^2 + k_1^2\right)\right] - \omega^2 \dfrac{Il}{Et} \end{bmatrix} \begin{Bmatrix} \hat{u}_1 \\ \hat{u}_2 \\ \hat{\phi} \end{Bmatrix} = \mathbf{0} \ , \qquad (50)$$

where the ratio $\varsigma = (t/l)^2$ has been introduced.

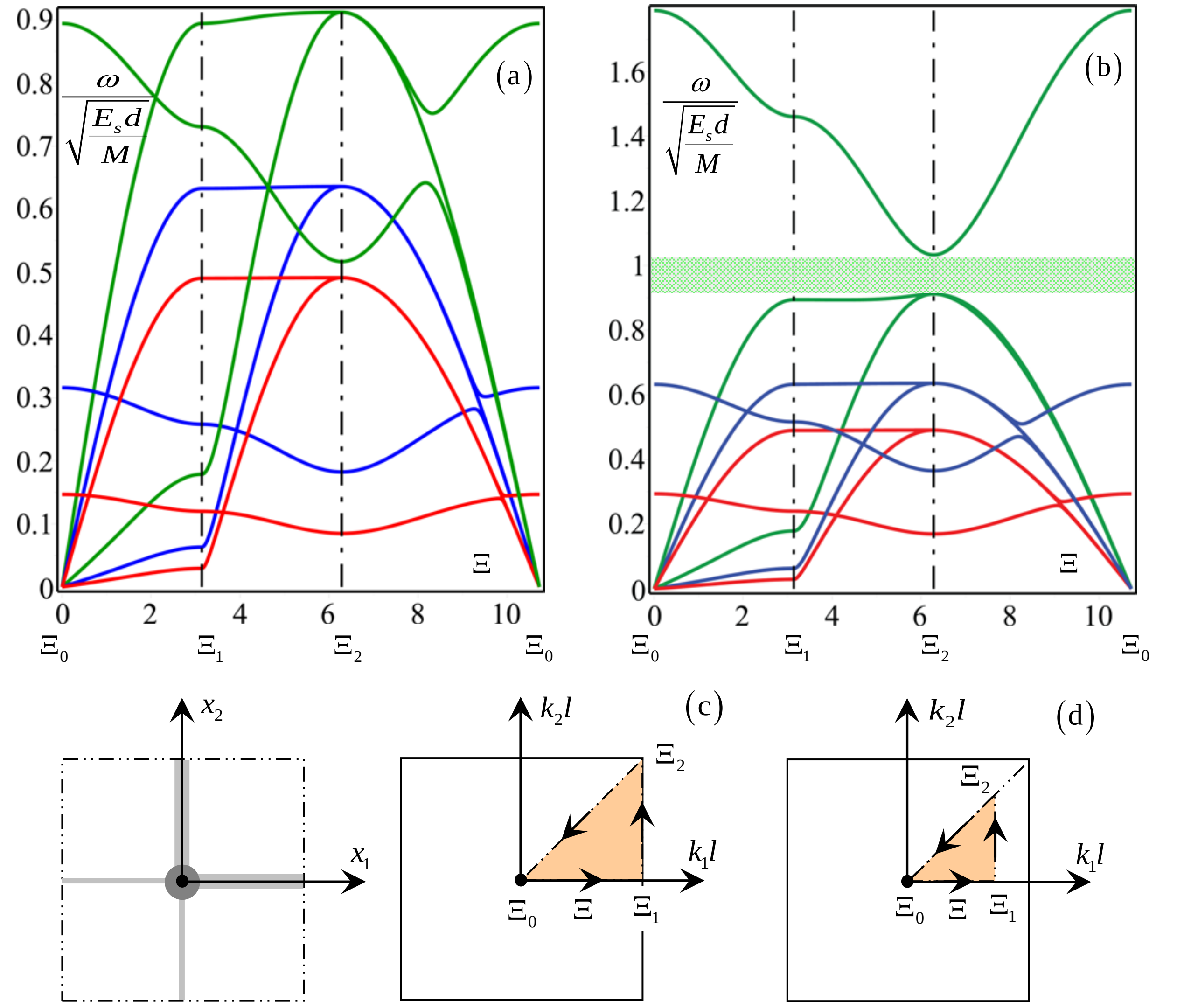

Figure 5: Influence of the ratio $t/l$ on the band structure of the centrosymmetric square lattice ($\alpha = 1$) along the closed polygonal curve $\Upsilon$ ($t/l = 3/50$ red; $t/l = 1/10$ blue; $t/l = 1/5$ green) for varying the rotational inertia of the nodes $J/Ml^2$: (a) $J/Ml^2 = 1/50$, (b) $J/Ml^2 = 1/200$; (c) Periodic cell and Brillouin zone (highlighted in orange the reduced Brillouin zone bounded by the curve $\Upsilon$); (d) Subdomain of the reduced Brillouin zone.

The dispersion function of the centrosymmetric lattice ($\alpha = 1$) obtained for the dimensionless moment of inertia $J/Ml^2 = 1/50$ and several values of the ratio $t/l$ are shown in Figure 5.a, where no stop band is observed. In these diagrams the two acoustic branches are clearly shown, together with the optical one starting from the critical point $\omega_{opt}\left(|\mathbf{k}| \to 0\right)/\sqrt{E_s d/M} \propto (t/l)^{\frac{3}{2}}$ at $\Xi_0$ ($\mathbf{k} = \mathbf{0}$) with vanishing group velocity $v_g = 0$. This critical frequency is increasing with the ratios $t/l$ and $\alpha$. Several crossing points between the optical and acoustic branches may be observed. For $t/l = 1/5$, in the range $\Xi \in \left[2\pi, \left(2+\sqrt{2}\right)\pi\right]$, a veering between the optical and the acoustic branches takes place, i.e. a repulsive phenomenon between the two branches. Moreover, for long wavelengths limit the optical branch is associated to rotational and transverse waves, while the propagation of pressure waves is obtained in the second acoustic branch.

When reducing the moment of inertia of the nodes $J/Ml^2 = 1/200$ the angular frequencies of the optical branches increase as shown by the diagrams in Figure 5.b and band gaps appear when decreasing the ligament thickness (see the case $t/l = 1/5$). Several crossing are detected between the acoustic and optical branches, while for $t/l = 1/10$ (blu line) a veering between the acoustic and optical branches is observed in the third range of the domain $\Upsilon$.

The reliability of the micropolar model obtained in Section 3 is analysed by comparing the dispersive functions with those from the discrete model. Being the micropolar model formulated for the long wavelength limit, the comparison is carried out in the subdomain of the reduced Brillouin zone shown in Figure 5.d. The dispersive functions in case of centrosymmetric lattice ($\alpha = 1$) and for $J/Ml^2 = 1/50$ are shown in Figure 6.a. for three values of the ratio $t/l$ for both the discrete model (continuous line) and the micropolar model (dashed line). This comparison shows the diagrams in Figure 6.a differ less than 5% for $|\mathbf{k}| l \le \sqrt{2}/3\ \pi$, i.e. $\lambda \ge 4.2l$, while for $|\mathbf{k}| l \le \sqrt{2}/2\ \pi$, i.e. $\lambda \ge 2.8l$, an error less than 10% is obtained. This accuracy is achieved because the hermitian matrix of the eigenproblem by the micropolar model is a second order approximation of the corresponding one by the discrete model. In fact, the comparison with the dispersion curves by the constitutive model derived assuming a first order expansion of the rotational field and providing the positive defined constitutive parameter $S^+ = \frac{(K_{11} + K_{22})}{2}$ is given in the diagrams of Figure 6.b. From this comparison, a reduced accuracy of the model to represent the optical branch is observed when increasing the wave number.

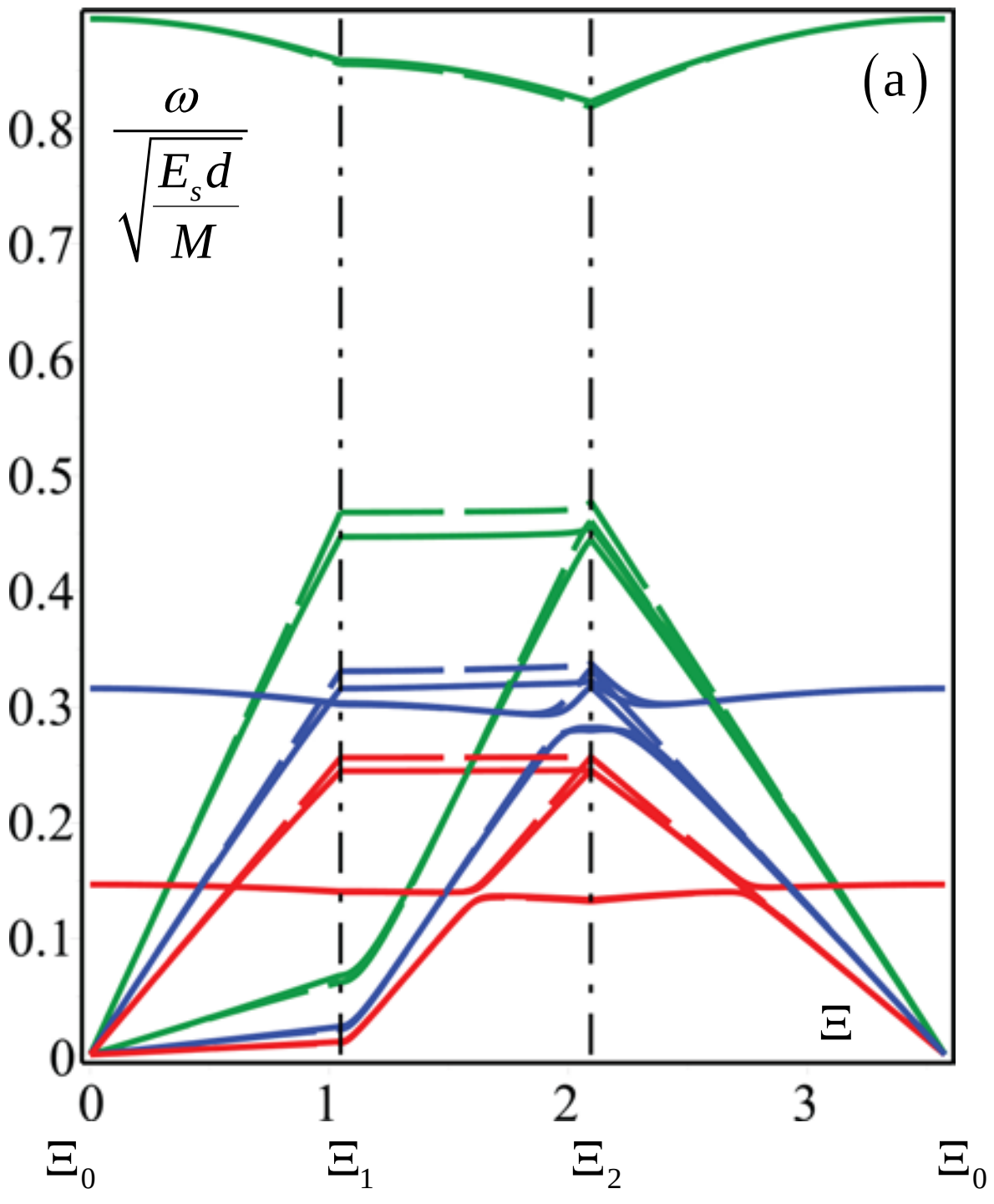

Figure 6: Dispersive functions for centrosymmetric square lattice ($\alpha = 1$, $J/Ml^2 = 1/50$). Comparison between the discrete model (continuous line) and the micropolar continuum model (dashed line) in a subdomain of the reduced Brillouin zone ($t/l = 15/250$ red; $t/l = 1/10$ blue; $t/l = 1/5$ green). (a) Constitutive constant $S$; (b) Constitutive constant $S^+$.

A more complete perception of the accuracy of the micropolar model may be given by the comparison between the discrete and the continuum model shown in Figure 7. Here waves propagating along axis $x_1$ ($k_2 = 0$) in a centrosymmetric lattice for a range of the dimensionless wave number $\Xi \in [0, \pi]$ ($\alpha = 1$, $J/Ml^2 = 1/50$, $t/l = 1/5$). In Figure 7.a the dashed black lines represent the band structure of the proposed continuum model ($S = -K_{12}$), which is in a good agreement with the theoretical one. In Figure 7.b the results are obtained from the positive definite model derived from a first order expansion of the rotational field. In this case, a poor accuracy appears in the optical branch.

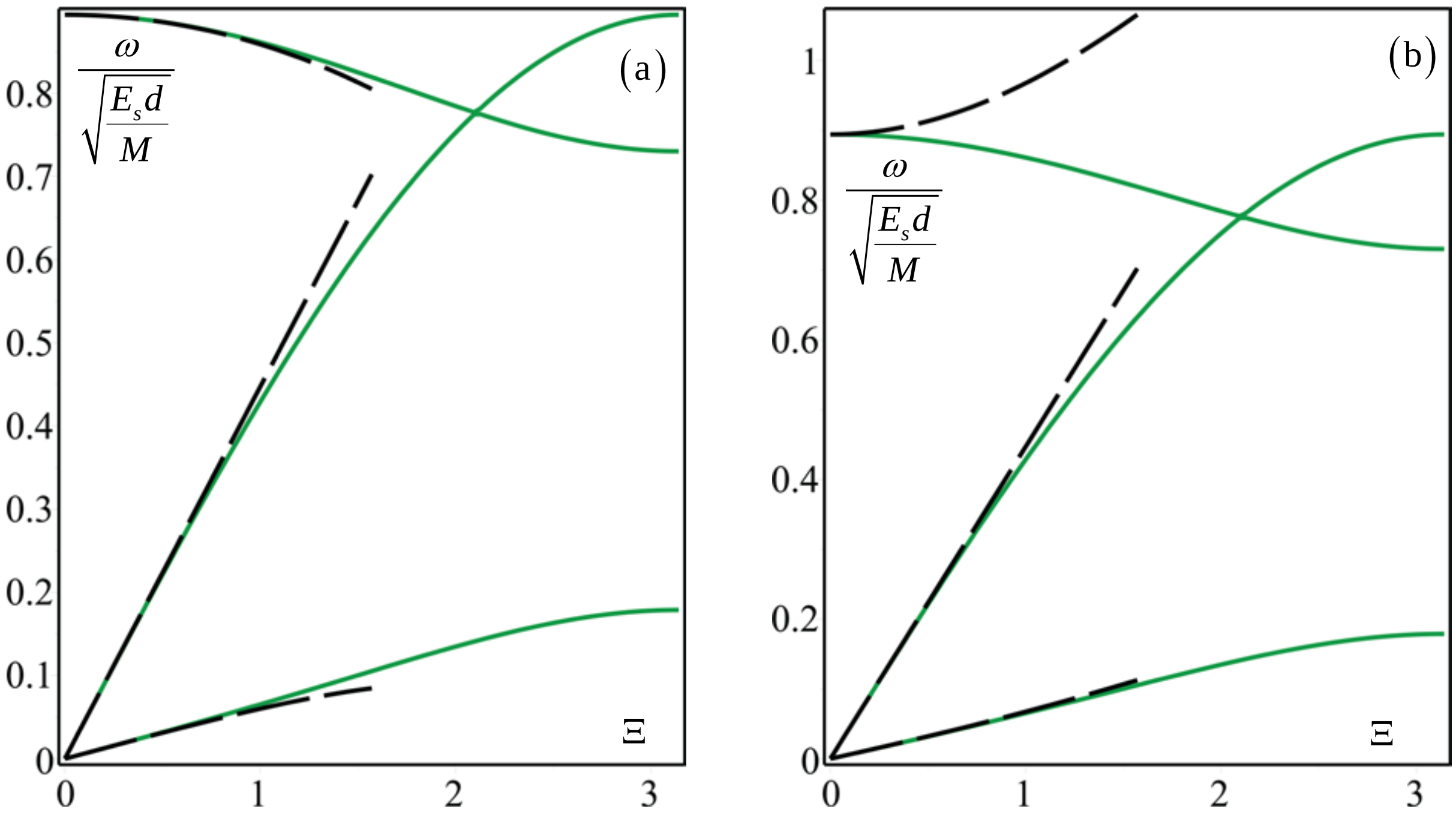

Figure 7: Dispersive functions for wave propagation along $x_1$ axis ($\alpha = 1$, $J/Ml^2 = 1/50$, $t/l = 1/5$). Comparison between the discrete model (continuous line) and the micropolar continuum model (dashed line). (a) Constitutive constant $S$; (b) Constitutive constant $S^+$.

The influence of the non-centrosymmetry on the band structure of the square lattice is shown in the diagrams of Figure 8 for different values of the stiffness step ratio $\alpha$. Increasing $\alpha$, a corresponding increase of the critical frequency of the optical branch in the long wavelength asymptotics is obtained due to an increase of the overall stiffness. The diagrams in Figure 8.a are obtained for the ratios $t/l = 15/250$ and $J/Ml^2 = 1/50$. Here the dispersion functions exhibit several crossing points between the optical and the two acoustic branches. Two veering points are obtained between the optical and the second acoustic branch in the domain $\Xi \in \left[\pi, \left(2+\sqrt{2}\right)\pi\right]$. Finally, it is worth to note that the band structure of the three considered beam-lattices does not exhibit stop bands. In Figure 8.b the diagrams corresponding to a small radius of gyration, namely $J/Ml^2 = 1/450$ are given. Here, higher values of the frequencies of the optical branches are obtained in comparison to Figure 8.a, according to the relation $\omega_{opt}\left(|\mathbf{k}| \to 0\right)/\sqrt{E_s d/M} \propto \left(J/Ml^2\right)^{-\frac{1}{2}}$. As a consequence, a low frequency stop-band (for wave vector scanning the curve $\Upsilon$) is observed for non-centrosymmetric lattice, whose amplitude increases with $\alpha$ .

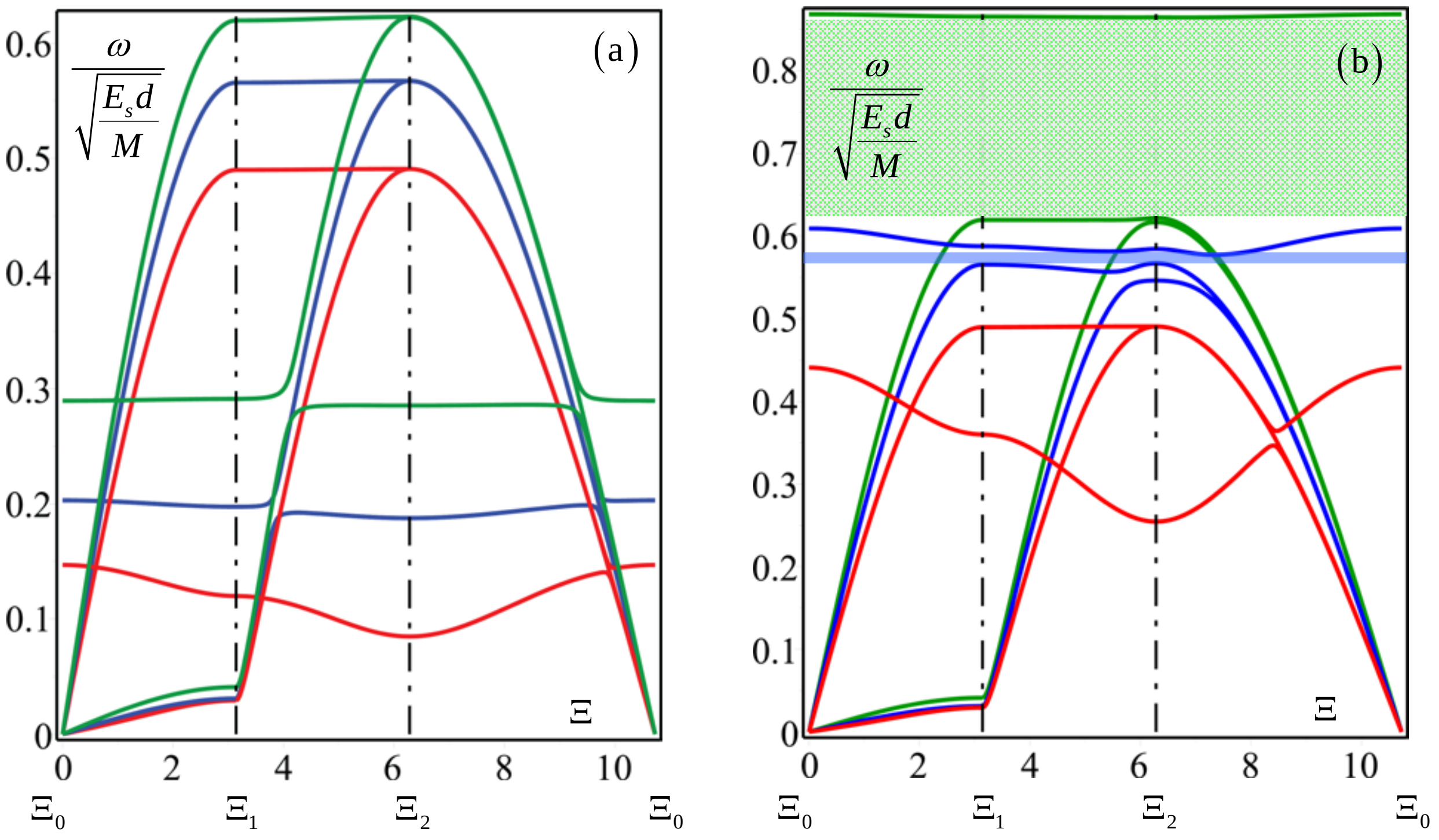

Figure 8: Band structure of the non-centrosymmetric square lattice along the closed polygonal curve $\Upsilon$ ($\alpha = 1$ red; $\alpha = 2$ blue; $\alpha = 4$ green) for dimensionless ligament thickness $t/l = 15/250$ and different dimensionless moment of inertia $J/Ml^2$ : (a) $J/Ml^2 = 1/50$, (b) $J/Ml^2 = 1/450$.

The accuracy of the micropolar model is represented in a synthetic way by the diagrams of Figure 9. Also in this case a good agreement between the discrete and the continuum model is shown in Figure 9.a for $|\mathbf{k}|l \leq \sqrt{2}/2\ \pi$, namely for wavelengths $\lambda \geq 2.8l$. Figure 9.b refers to the case of micropolar continuum based on a first order expansion of the rotational field, i.e. for $S^+ = \frac{(K_{11} + K_{22})}{2}$. From these diagrams the poor accuracy of the optical branch may be observed, while the acoustic branches are equally well approximated in the subdomain of the reduced Brillouin zones.

In consideration of the negative definiteness of the elastic constant $S = -K_{12}$ of the micropolar model, the hyperbolicity of the equation of motion has to be ensured. This circumstance is verified referring to the Legendre–Hadamard ellipticity conditions (semi-ellipticity) requiring real values of the wave velocity, namely the positivity of the square of the dispersion function (see for reference Jeong and Neff, 2010 and Eremeyev *et al.*, 2013). Since unconditional hyperbolicity cannot be ensured, some meaningful cases have been considered by controlling the positivity of the square of the dispersion functions in the Brillouin zone (see Figure 5.c). In Figure 10 the square of the dispersive surfaces for the following non dimensionless parameters $t/l = 1/5$, $J/Ml^2 = 1/50$ and $\alpha = 1$ are plotted. It may be noted that both the second acoustic surface and the optical surface are positive in the whole Brillouin zone, while the lower acoustic surface turns out to be positive in the red domain shown in Figure 10.b. However, it must be noted that negative values are attained at points for which $|\mathbf{k}| l \geq \sqrt{2}/2\ \pi$, i.e. where the continuum micropolar model loses accuracy. The case of non-centrosymmetric lattice $\alpha = 4$ is shown in Figure 11, where a qualitative behavior analogous to the centrosymmetric case is observed, together with a reduction of the regions where the square of the dispersive function is negative.

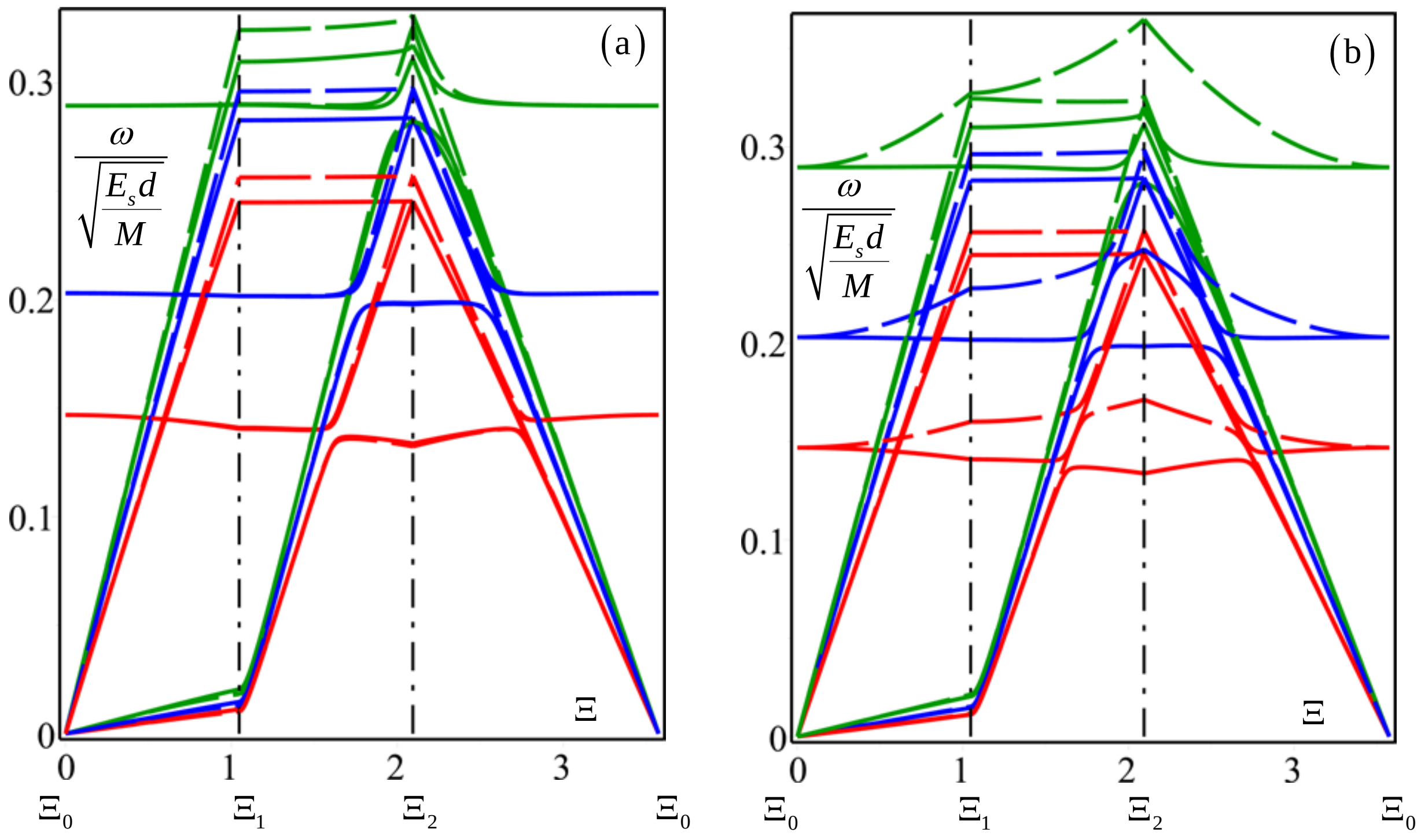

Figure 9: Dispersive functions for square lattice ($t/l = 15/250$, $J/Ml^2 = 1/50$). Comparison between the discrete model (continuous line) and the micropolar continuum model (dashed line) in a subdomain of the reduced Brillouin zone ($\alpha = 1$ red; $\alpha = 2$ blue; $\alpha = 4$ green). (a) Constitutive constant $S$; (b) Constitutive constant $S^+$.

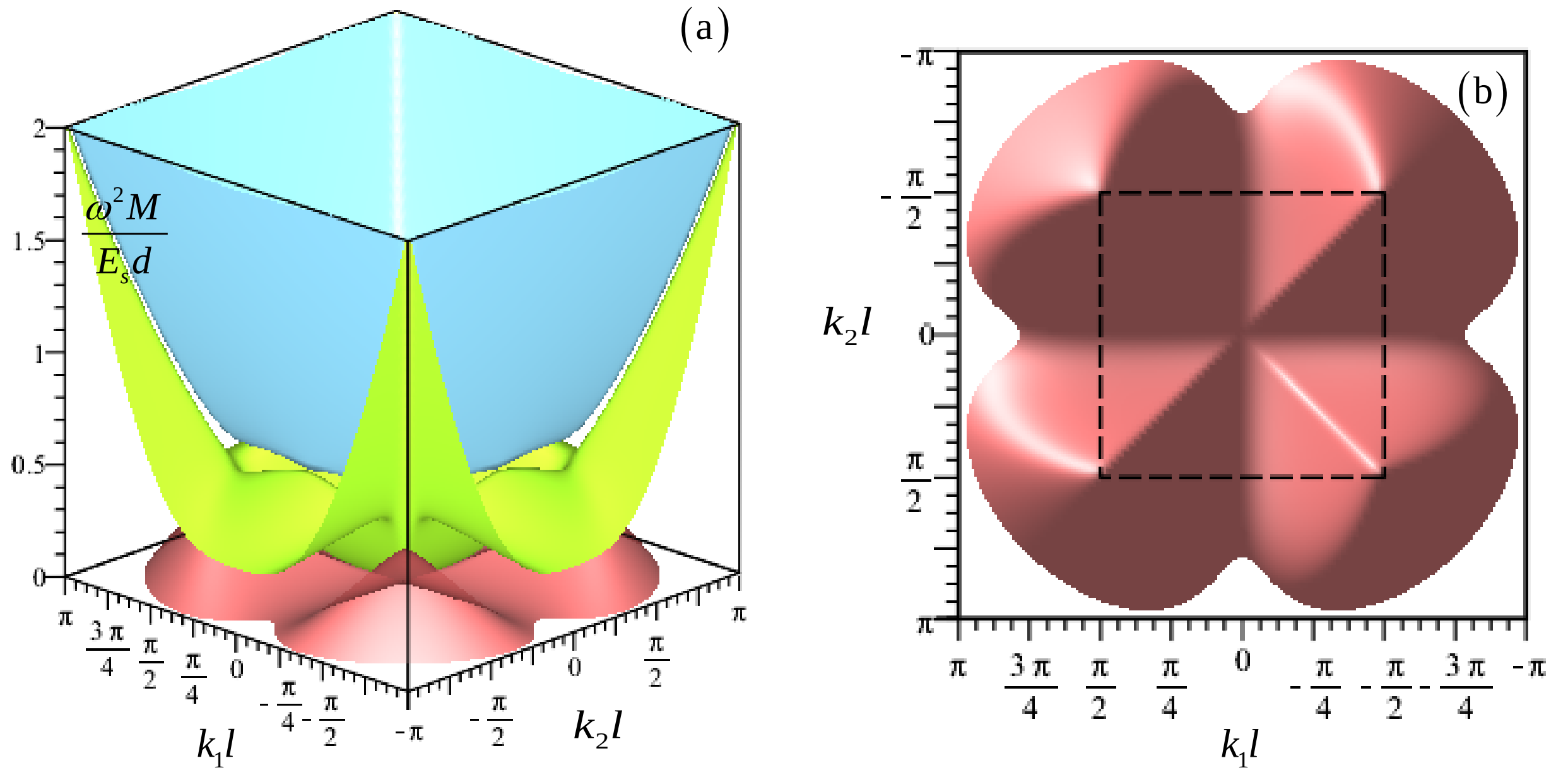

Figure 10: Square of the dispersive surfaces in the Brillouin zone by the micropolar continuum model for centrosymmetric lattice with $t/l = 1/5$, $J/Ml^2 = 1/50$, $\alpha = 1$. (a) Positive Floquet-Bloch spectrum; (b) Domain of positivity of the first acoustic surface and square subdomain of good accuracy of the model.

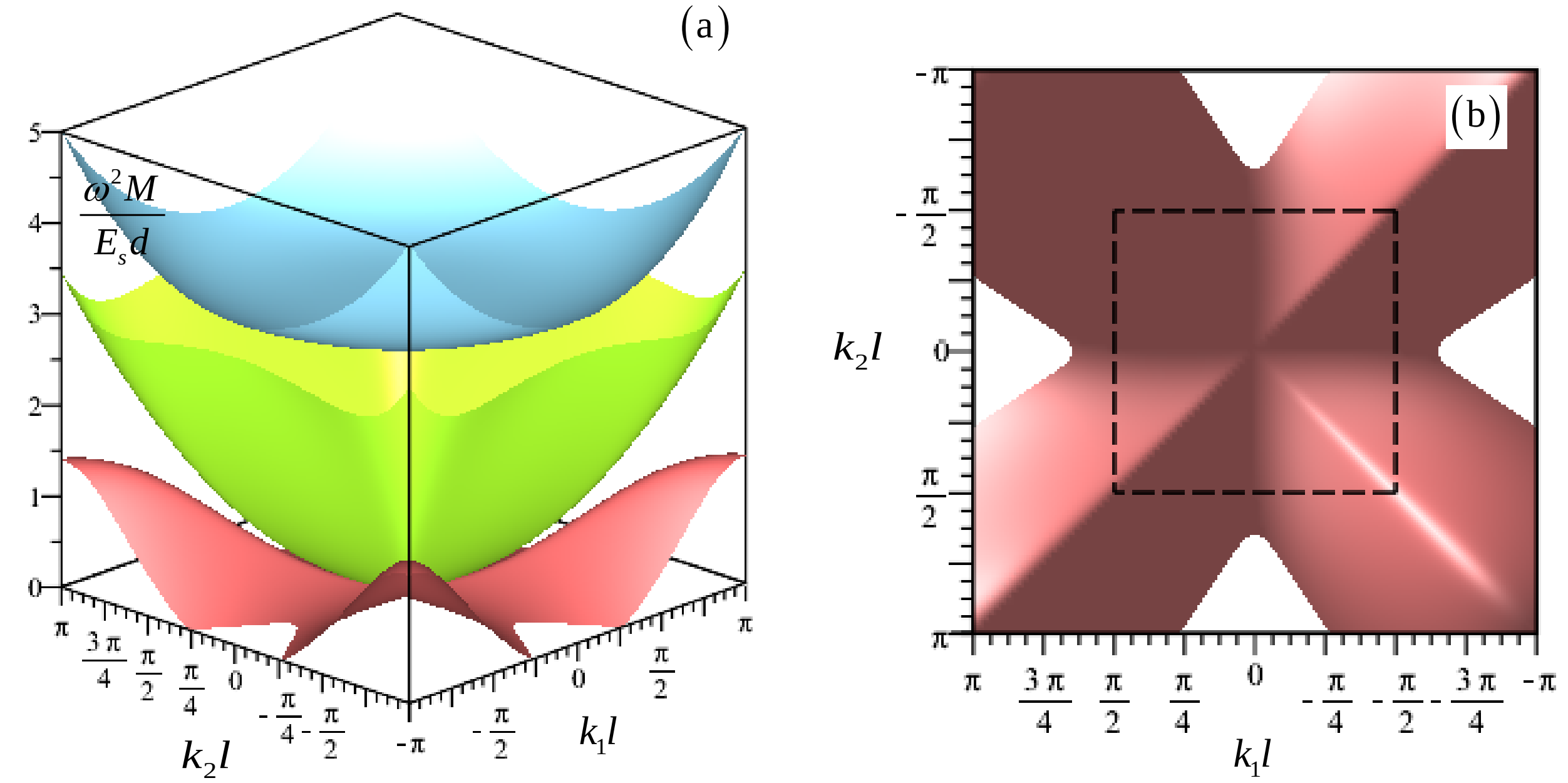

Figure 11: Square of the dispersive surfaces in the Brillouin zone by the micropolar continuum model for non-centrosymmetric lattice with $t/l = 1/5$, $J/Ml^2 = 1/50$, $\alpha = 4$. (a) Positive Floquet-Bloch spectrum; (b) Domain of positivity of the first acoustic surface and square subdomain of good accuracy of the model.

*5.2. Equilateral triangular lattice*

The equilateral triangular lattice and the hexagonal periodic cell are shown in Figure 12. The system is hexagonal with the relevant geometrical data are $A_{cell} = \sqrt{3}l^2$, $\mathbf{n}_1 = \mathbf{e}_1$, $\mathbf{t}_1 = \mathbf{e}_2$, $\mathbf{n}_2 = \frac{1}{2}\mathbf{e}_1 + \frac{\sqrt{3}}{2}\mathbf{e}_2, \mathbf{t}_2 = -\frac{\sqrt{3}}{2}\mathbf{e}_1 + \frac{1}{2}\mathbf{e}_2$, $\mathbf{n}_3 = -\frac{1}{2}\mathbf{e}_1 + \frac{\sqrt{3}}{2}\mathbf{e}_2, \mathbf{t}_3 = -\frac{\sqrt{3}}{2}\mathbf{e}_1 - \frac{1}{2}\mathbf{e}_2$. Because the assumed geometry, the ligament stiffnesses have the following properties $K_n^1 = K_n^2 = K_n^3$, $K_{11}^1 = K_{22}^2 = K_{11}^3$ , $K_{22}^1 = K_{11}^2 = K_{22}^3$, $K_{12}^1 = K_{12}^2 = K_{12}^3$, so that the stiffness terms are $\hat{K} = \hat{K}^1 = \hat{K}^2 = \hat{K}^3 = K_{11}^i + 2K_{12}^i + K_{22}^i$ e $\tilde{K} = \tilde{K}^1 = -\tilde{K}^2 = \tilde{K}^3 = \left(K_{11}^i - K_{22}^i\right)/2$.

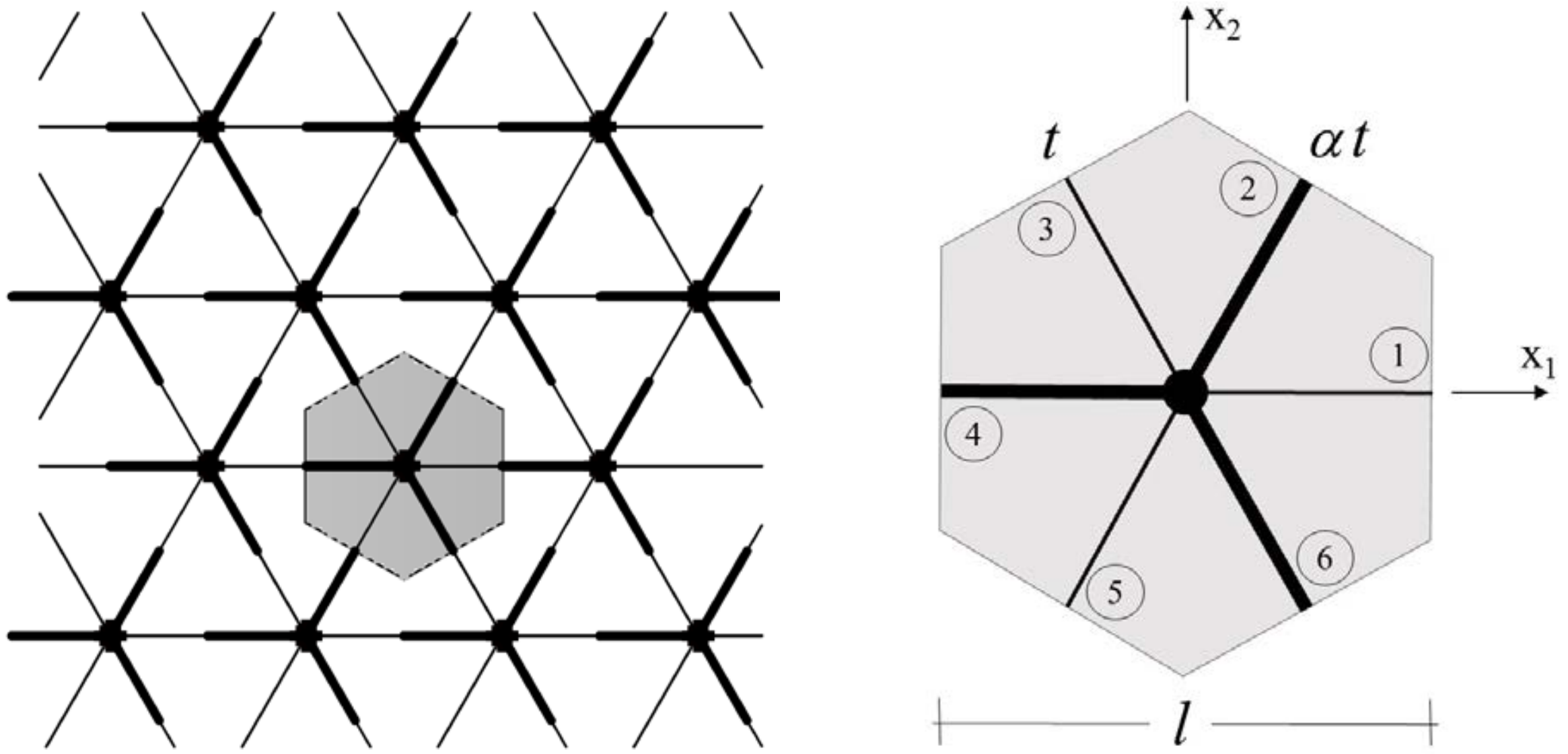

(a) (b)

Figure 12: Periodic non-centrosymmetric equilateral triangular lattice.

The eigenvalue problem (11) ruling the harmonic plane wave propagation in the Lagrangian model takes the form:

$$\begin{bmatrix} \frac{1}{2}K_n \mathrm{f}_1 + \frac{1}{2}\left(K_n + \frac{3\hat{K}}{l^2}\right)(\mathrm{f}_2 + \mathrm{f}_3) - \omega^2 M & \frac{\sqrt{3}}{2}\left(K_n - \frac{\hat{K}}{l^2}\right)\mathrm{f}_2 - \mathrm{f}_3 & \frac{\sqrt{3}}{2}\frac{\tilde{K}}{l}\left[(\mathrm{f}_2 - \mathrm{f}_3) + i(g_2 + \mathrm{g}_3)\right] \\ \frac{\sqrt{3}}{2}\left(K_n - \frac{\hat{K}}{l^2}\right)(\mathrm{f}_2 - \mathrm{f}_3) & \frac{1}{2}\left(3K_n + \frac{\hat{K}}{l^2}\right)(\mathrm{f}_2 + \mathrm{f}_3) - \omega^2 M & \frac{\tilde{K}}{2l}\left[(2\mathrm{f}_1 - \mathrm{f}_2 - \mathrm{f}) + i(2g_1 + g_2 - g_3)\right] \\ \frac{\sqrt{3}}{2}\frac{\tilde{K}}{l}\left[(\mathrm{f}_2 - \mathrm{f}_3) - i(g_2 + \mathrm{g}_3)\right] & \frac{\tilde{K}}{2l}\left[(2\mathrm{f}_1 - \mathrm{f}_2 - \mathrm{f}) - i(2g_1 + g_2 - g_3)\right] & \left[3\hat{K} - 2K_{12}(\mathrm{f}_1 + \mathrm{f}_2 + \mathrm{f}_3)\right] - \omega^2 J \end{bmatrix} \begin{Bmatrix} \hat{u}_1 \\ \hat{u}_2 \\ \hat{\phi} \end{Bmatrix} = \mathbf{0}, \quad (51)$$

having defined $f_i = f_i(\mathbf{k},\mathbf{n}_i) = 1-\cos(l\mathbf{k}\cdot\mathbf{n}_i)$ and $g_i = g_i(\mathbf{k},\mathbf{n}_i) = \sin(l\mathbf{k}\cdot\mathbf{n}_i)$, $i$=1,3. When considering the long wave approximation, the constitutive equation of the micropolar continuum is derived from (21), (22) and (28) and takes the form

$$\begin{Bmatrix} \sigma_{11} \\ \sigma_{22} \\ \sigma_{12} \\ \sigma_{21} \\ m_1 \\ m_2 \end{Bmatrix} = \begin{bmatrix} 2\mu+\lambda & \lambda & 0 & 0 & 0 & Y \\ \lambda & 2\mu+\lambda & 0 & 0 & 0 & -Y \\ 0 & 0 & \mu+\kappa & \mu-\kappa & Y & 0 \\ 0 & 0 & \mu-\kappa & \mu+\kappa & Y & 0 \\ 0 & 0 & Y & Y & S & 0 \\ Y & -Y & 0 & 0 & 0 & S \end{bmatrix} \begin{Bmatrix} \gamma_{11} \\ \gamma_{22} \\ \gamma_{12} \\ \gamma_{21} \\ \chi_1 \\ \chi_2 \end{Bmatrix} , \qquad (52)$$

with five elastic moduli

$$\mu = \sqrt{3}\frac{K_n l^2 + \hat{K}}{4l^2}, \qquad \lambda = \sqrt{3}\frac{K_n l^2 - \hat{K}}{4l^2}, \qquad \kappa = \sqrt{3}\frac{\hat{K}}{2l^2} ,$$
$$Y = \sqrt{3}\frac{\tilde{K}}{2l^2}, \qquad S = -\sqrt{3}K_{12} . \qquad (53)$$

When only the first order description of the rotational field is considered, the elastic modulus involving the curvatures takes the positive defined form $S^+ = \frac{\sqrt{3}}{2}(K_{11}+K_{22})$. In case of symmetric macro-strain fields $\gamma_{12} = \gamma_{21}$ with vanishing curvature $\boldsymbol{\chi} = \mathbf{0}$, the elastic moduli of a classical continuum are obtained $C_{1111} = 2\mu+\lambda$, $C_{1122} = \lambda$ and $C_{1212} = \mu$ characterizing a transversely isotropic material. The equation providing the dispersion function in the equivalent micropolar continuum is written as

$$\begin{bmatrix} (2\mu+\lambda)k_1^2 + (\mu+\kappa)k_2^2 - \rho\omega^2 & (\mu-\kappa+\lambda)k_1 k_2 & 2Yk_1 k_2 - 2\kappa i k_2 \\ (\mu-\kappa+\lambda)k_1 k_2 & (2\mu+\lambda)k_2^2 + (\mu+\kappa)k_1^2 - \rho\omega^2 & Y(k_1^2 - k_2^2) + 2\kappa i k_1 \\ 2Yk_1 k_2 + 2\kappa i k_2 & Y(k_1^2 - k_2^2) - 2\kappa i k_1 & 4\kappa + S(k_1^2 + k_2^2) - \omega^2 I \end{bmatrix} \begin{Bmatrix} \hat{v}_1 \\ \hat{v}_2 \\ \hat{\theta} \end{Bmatrix} = \mathbf{0} . \qquad (54)$$

As for the previous example, in the long-wave limit $\lambda \to \infty$, the angular frequencies are $\omega_{aco1,2} = 0$ and $\omega_{opt} = \sqrt{\frac{2\kappa}{I}}$, the third one being depending on the non-centrosymmetry of the periodic cell being $\kappa = \frac{\sqrt{3}}{2}\frac{K_{11}^i + 2K_{12}^i + K_{22}^i}{l^2}$. Finally, in case of centrosymmetric cell, i.e. $\alpha = 1$ ,

the following elastic moduli are obtained $\mu = \frac{\sqrt{3}}{4} E \frac{t}{l}\left[1+\left(\frac{t}{l}\right)^2\right]$, $\lambda = \frac{\sqrt{3}}{4} E \frac{t}{l}\left[1-\left(\frac{t}{l}\right)^2\right]$, $\kappa = \frac{\sqrt{3}}{2} E \left(\frac{t}{l}\right)^3$, $Y = 0$, $S = -\frac{\sqrt{3}}{6} E l^2 \left(\frac{t}{l}\right)^3$ and the eigenproblem takes the form

$$\begin{bmatrix} (3+\varsigma)k_1^2 + (1+3\varsigma)k_2^2 - \frac{4\sqrt{3}l\rho}{3Et}\omega^2 & 2(1-\varsigma)k_1 k_2 & -4i\varsigma k_2 \\ 2(1-\varsigma)k_1 k_2 & (1+3\varsigma)k_1^2 + (3+\varsigma)k_2^2 - \frac{4\sqrt{3}l\rho}{3Et}\omega^2 & 4i\left(\frac{t}{l}\right)^2 k_1 \\ 4i\varsigma k_2 & -4i\varsigma k_1 & \frac{2}{3}\varsigma\left[12 - (k_1 l)^2 - (k_2 l)^2\right] - \frac{4\sqrt{3}Il}{3Et}\omega^2 \end{bmatrix} \begin{Bmatrix} \hat{u}_1 \\ \hat{u}_2 \\ \hat{\varphi} \end{Bmatrix} = \mathbf{0}\,, \qquad (55)$$

being $\varsigma = (t/l)^2$.

The dispersion function of the centrosymmetric lattice ($\alpha = 1$) obtained for the dimensionless moment of inertia $J/Ml^2 = 1/50$ and several values of the ratio $t/l$ are shown in Figure 13.a, where no stop band is observed. Here two acoustic branches are clearly shown, together with the optical one characterized by a critical point for long wavelength limit, in qualitative agreement with the results obtained for square lattice. Several crossing points between the optical and acoustic branches may be observed. For $t/l = 1/10$ and $t/l = 1/5$, a veering between the optical and the acoustic branches is observed in the domains $\Xi \in [0, 4/3\pi]$ and $\Xi \in \left[2\pi, 2\pi\left(1+\sqrt{3}/3\right)\right]$ along the curve $\Upsilon$. In this case, no band gap is obtained. For lower values of the dimensionless moment of inertia of the nodes, i.e. $J/Ml^2 = 1/200$, the dispersion functions are plotted in Figure 13.b, where a band gap is shown for the ratio $t/l = 1/5$. Several points of crossing may be observed between the acoustic branches and the optical one. In particular, for $t/l = 1/10$ a veering point is observed between the second acoustic branch and the optical one.

The diagrams in Figure 14.a, where the dispersive functions obtained from the discrete and the micropolar models of a centrosymmetric lattice with dimensionless moment of inertia $J/Ml^2 = 1/50$ may be compared, allow assessing the accuracy of the equivalent micropolar model. This comparison shows the diagrams in Figure 14.a differ less than $5\%$ for $|\mathbf{k}|l \leq 4/9\ \pi$, i.e. $\lambda \geq 4.5l$, while for $|\mathbf{k}|l \leq 2/3\ \pi$, i.e. $\lambda \geq 3l$, an error less than $10\%$ is obtained. The same comparison is shown by the diagrams in Figure 14.b with reference to the micropolar model obtained as in Section 3.2 by assuming a first order description of the rotational field, i.e. by assuming the definite positive elastic modulus $S^{+}=\frac{\sqrt{3}}{2}\left(K_{11}+K_{22}\right)$. As already noticed in the previous example, this model turns out with a limited accuracy in approximating the optical branch.

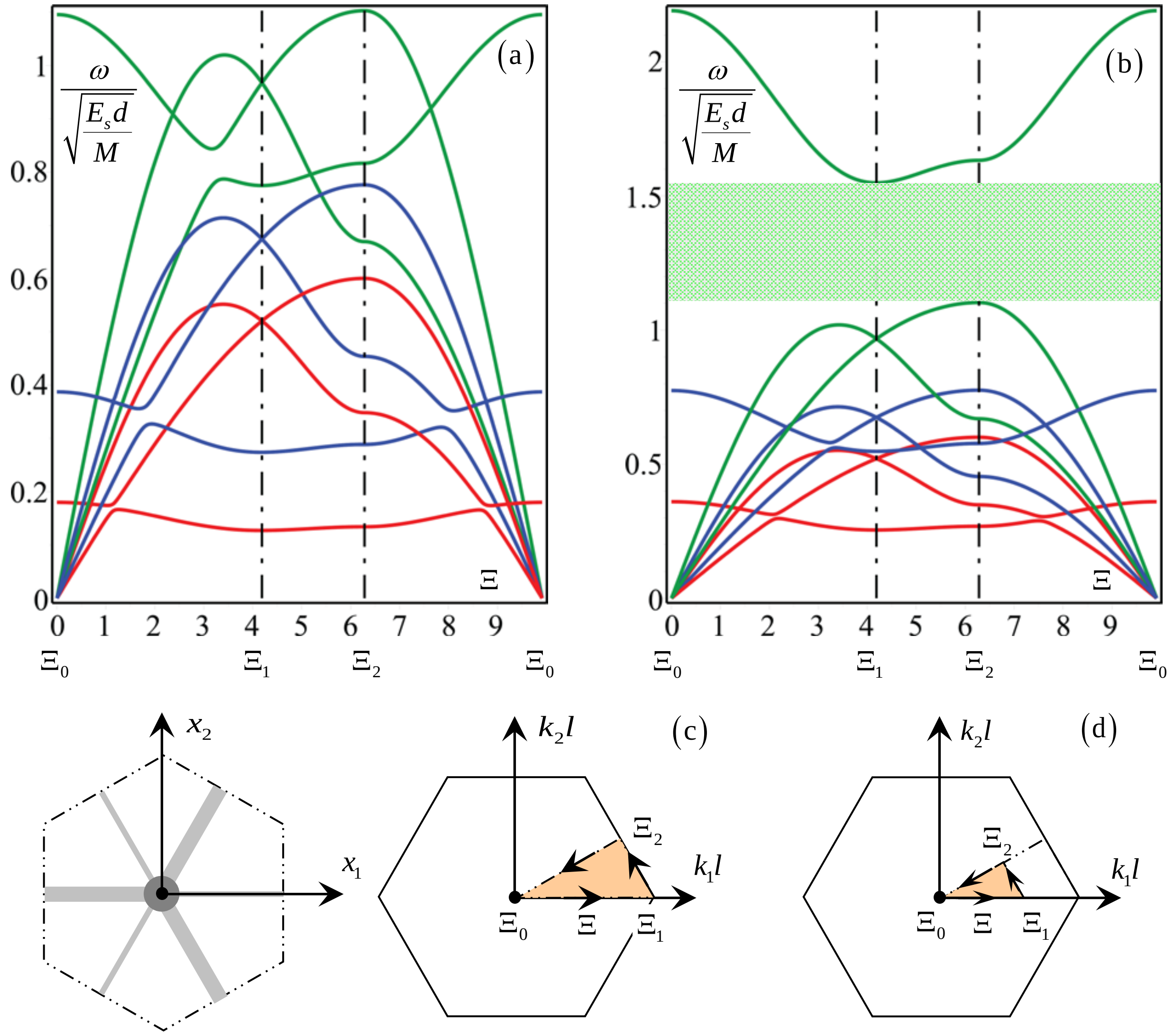

Figure 13: Influence of the dimensionless ligament thickness $t/l$ on the band structure of the equilateral triangular lattice ($\alpha=1$) along the closed polygonal curve $\Upsilon$ ($t/l=3/50$ red; $t/l=1/10$ blue; $t/l=1/5$ green) for different dimensionless mass moment of inertia $J/Ml^2$: (a) $J/Ml^2=1/50$, (b) $J/Ml^2=1/200$; (c) Periodic cell and Brillouin zone (highlighted in orange the reduced Brillouin zone bounded by the curve $\Upsilon$); (d) Subdomain of the reduced Brillouin zone.

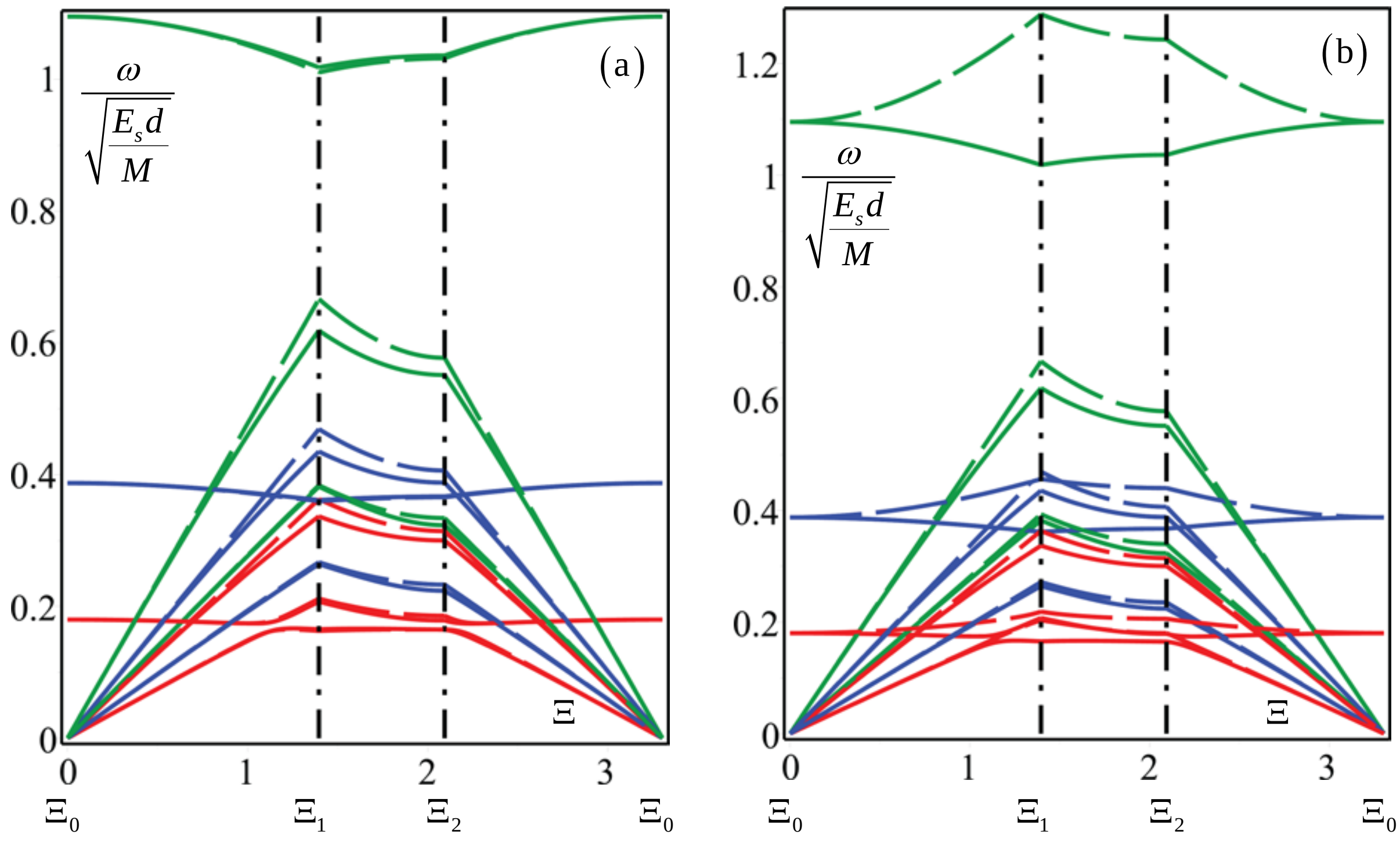

Figure 14: Dispersive functions for equilateral triangular lattice ($\alpha = 1$, $J/Ml^2 = 1/50$). Comparison between the discrete model (continuous line) and the micropolar continuum model (dashed line) in a subdomain of the reduced Brillouin zone ($t/l = 15/250$ red; $t/l = 1/10$ blue; $t/l = 1/5$ green). (a) Constitutive constant $S$; (b) Constitutive constant $S^+$.

A comparison of the accuracy of the two micropolar models is given in the diagrams of Figure 15 for the case of wave propagation along the $x_1$ axis, i.e. $k_2 l = 0$ and $k_1 l = \Xi \in [0, \pi]$, for a lattice with $\alpha = 1$, $J/Ml^2 = 1/50$, $t/l = 3/50$. While in the diagrams of Figure 15.a an excellent accuracy is obtained with the micropolar model here presented, with a good simulation of the discrete model in presence of veering points, a poor accuracy (see Figure 15.b) characterizes the micropolar model having a positive defined elastic modulus $S^+ = \frac{\sqrt{3}}{2}(K_{11} + K_{22})$.

The dispersion functions obtained from lattices with non-centrosymmetric topology are shown in the diagrams of Figure 16 for different values of the stiffness step ratio $\alpha$ and assuming dimensionless ligament thickness $t/l = 3/50$. As highlighted in the previous example, the critical point in the optical branch is characterized by a frequency that is increasing with the ratio $\alpha$, namely with the departure from the centrosymmetry of the periodic cell. Decreasing the dimensionless moment of inertia from $J/Ml^2 = 1/50$ (Figure 16.a) to $J/Ml^2 = 1/450$ (Figure 16.b) higher values of the frequency of the critical point on the acoustic branch are obtained together with a band gap (for wave vector scanning the curve $\Upsilon$) between the acoustic and the optical branches when increasing the ratio $\alpha$.

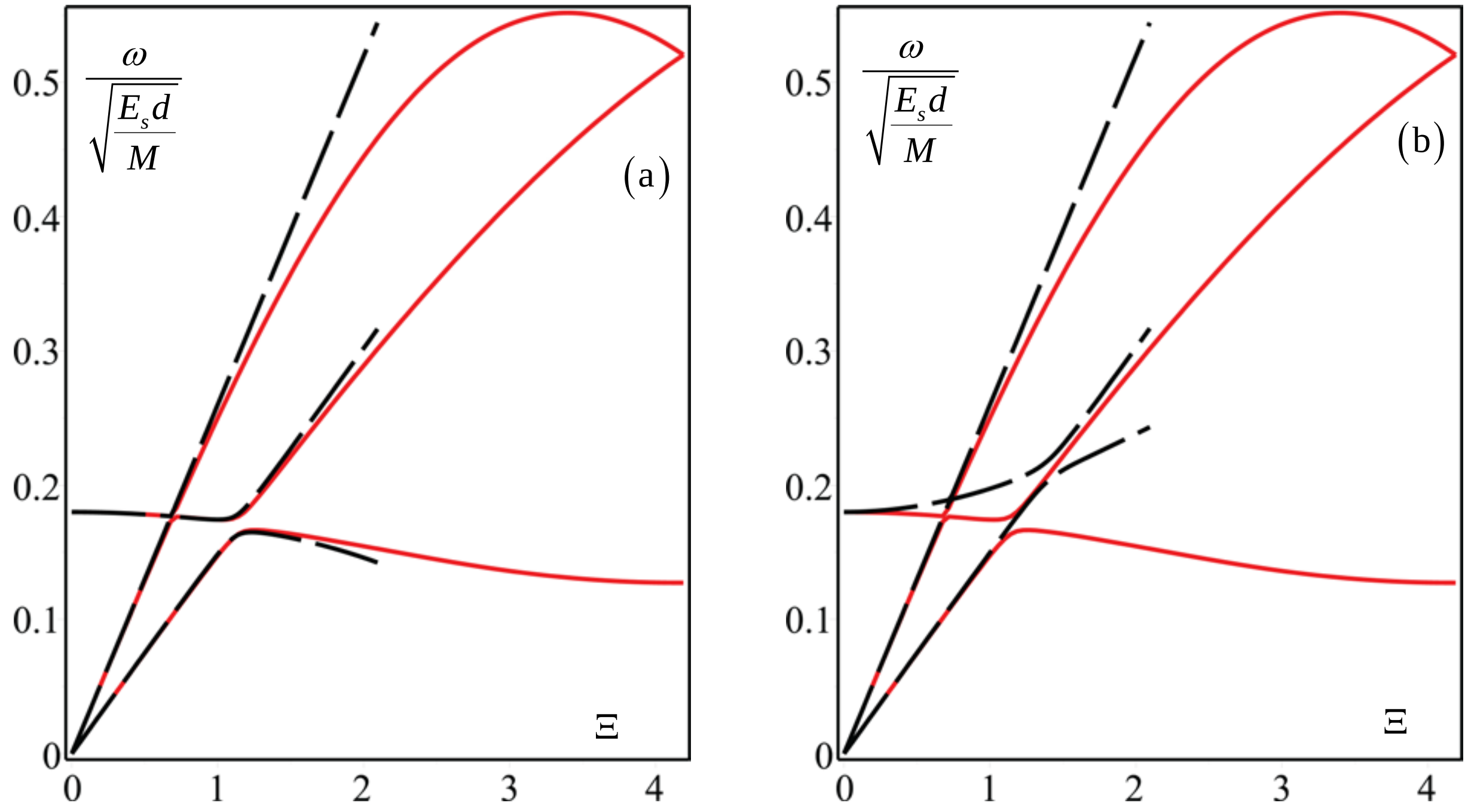

Figure 15: Dispersive functions for wave propagation along $x_1$ axis ($\alpha = 1$, $J/Ml^2 = 1/50$, $t/l = 3/50$). Comparison between the discrete model (continuous line) and the micropolar continuum model (dashed line). (a) Constitutive constant $S$; (b) Constitutive constant $S^+$.

The accuracy of the micropolar model is represented in a synthetic way by the diagrams of Figure 17. Also in this case a good agreement between the discrete and the continuum model is shown in Figure 17.a for $|\mathbf{k}| l \leq 2/3\ \pi$, namely for wavelengths $\lambda \geq 3l$. Figure 17.b refers to the case of micropolar continuum based on a first order expansion of the rotational field, i.e. for $S^+ = \frac{\sqrt{3}}{2}(K_{11} + K_{22})$. From these diagrams, the poor accuracy of the optical branch may be observed, while the acoustic branches are equally well approximated in the subdomain of the reduced Brillouin zones.

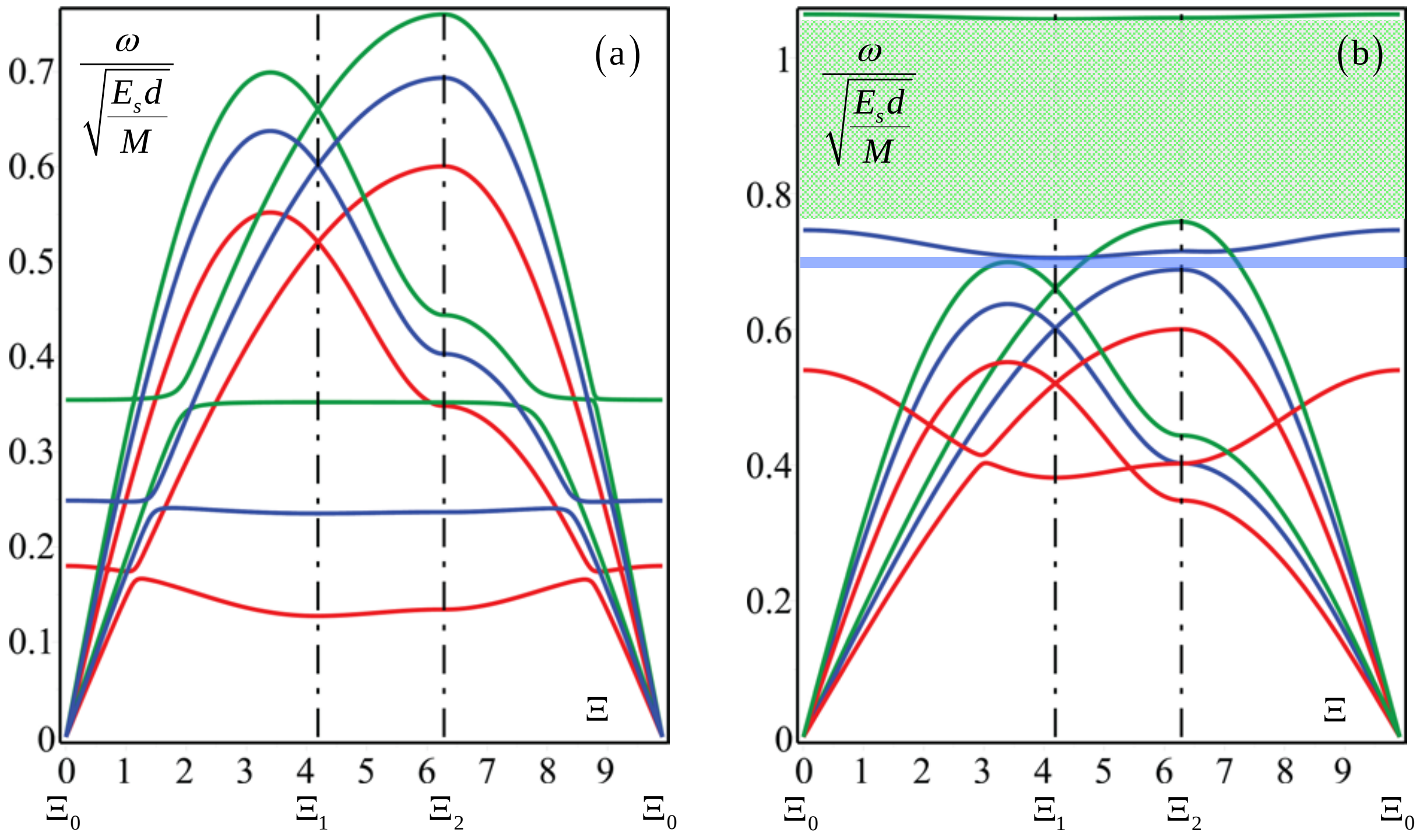

Figure 16: Band structure of the non-centrosymmetric equilateral triangular lattice along the closed polygonal curve $\Upsilon$ ($\alpha = 1$ red; $\alpha = 2$ blue; $\alpha = 4$ green) for dimensionless ligament thickness $t/l = 3/50$ and different dimensionless mass moment of inertia $J/Ml^2$ :

(a) $J/Ml^2 = 1/50$, (b) $J/Ml^2 = 1/450$.

The Legendre–Hadamard ellipticity conditions is here checked as explained in the previous example. In Figure 18 the surfaces representing the square of the dispersion function for a centrosymmetric lattice are plotted. From these diagrams, only the first acoustic surface attains negative values. In Figure 18.b the sub-region where this function is positive is the red one in the Brillouin zone. Also in this case it must be noted that negative values are attained at points $|\mathbf{k}| l \geq 2\sqrt{2}/3\ \pi$, i.e. where the micropolar model loses accuracy.

Finally, the case of non-centrosymmetric lattice $\alpha = 4$ is shown in Figure 19, where a qualitative behavior analogous to the centrosymmetric case is observed, together with a reduction of the regions where the square of the dispersive function is negative.

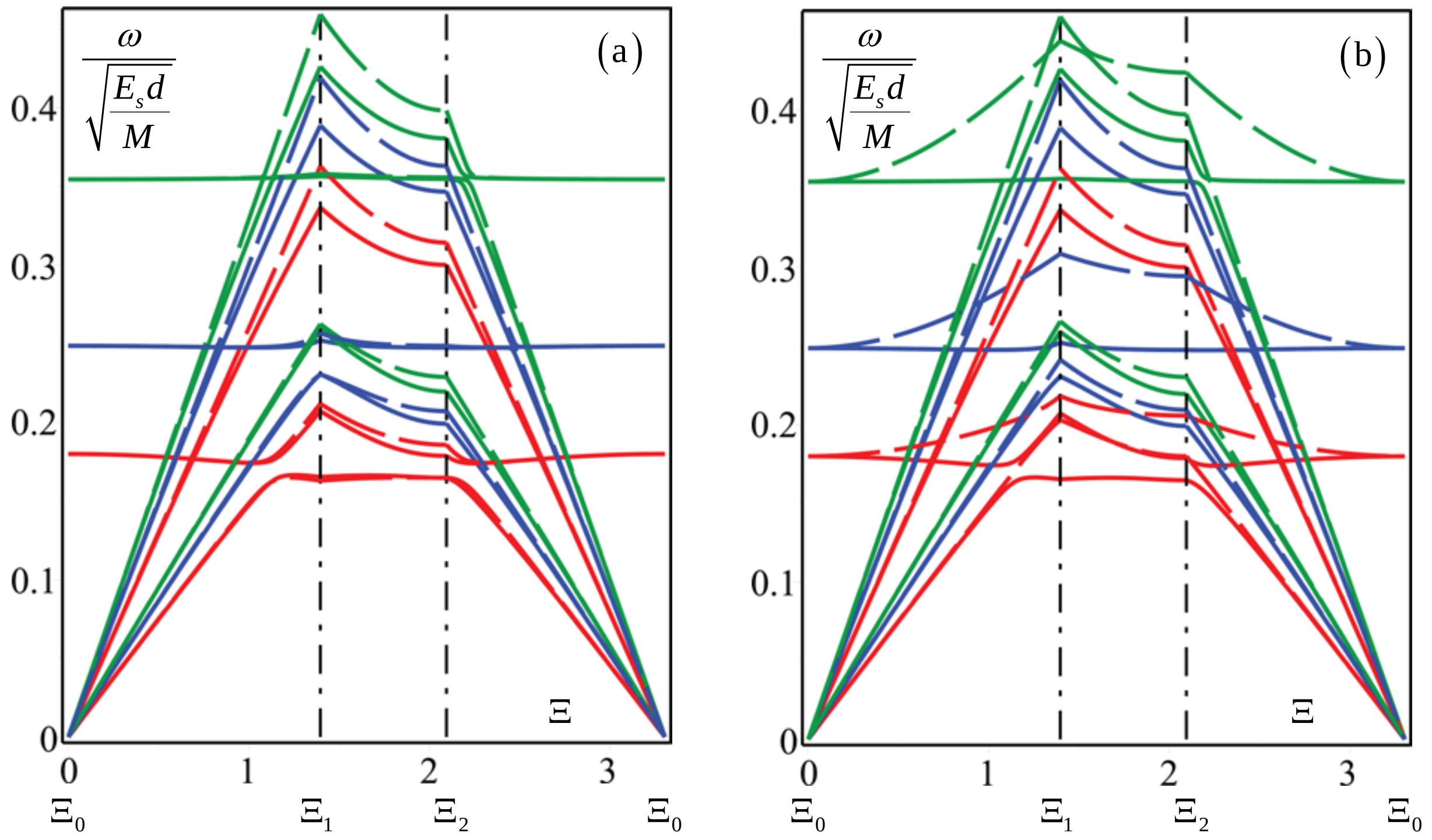

Figure 17: Dispersive functions for equilateral triangular lattice ($t/l = 3/50$, $J/Ml^2 = 1/50$). Comparison between the discrete model (continuous line) and the micropolar continuum model (dashed line) in a subdomain of the reduced Brillouin zone ($\alpha = 1$ red; $\alpha = 2$ blue; $\alpha = 4$ green). (a) Constitutive constant $S$; (b) Constitutive constant $S^+$.

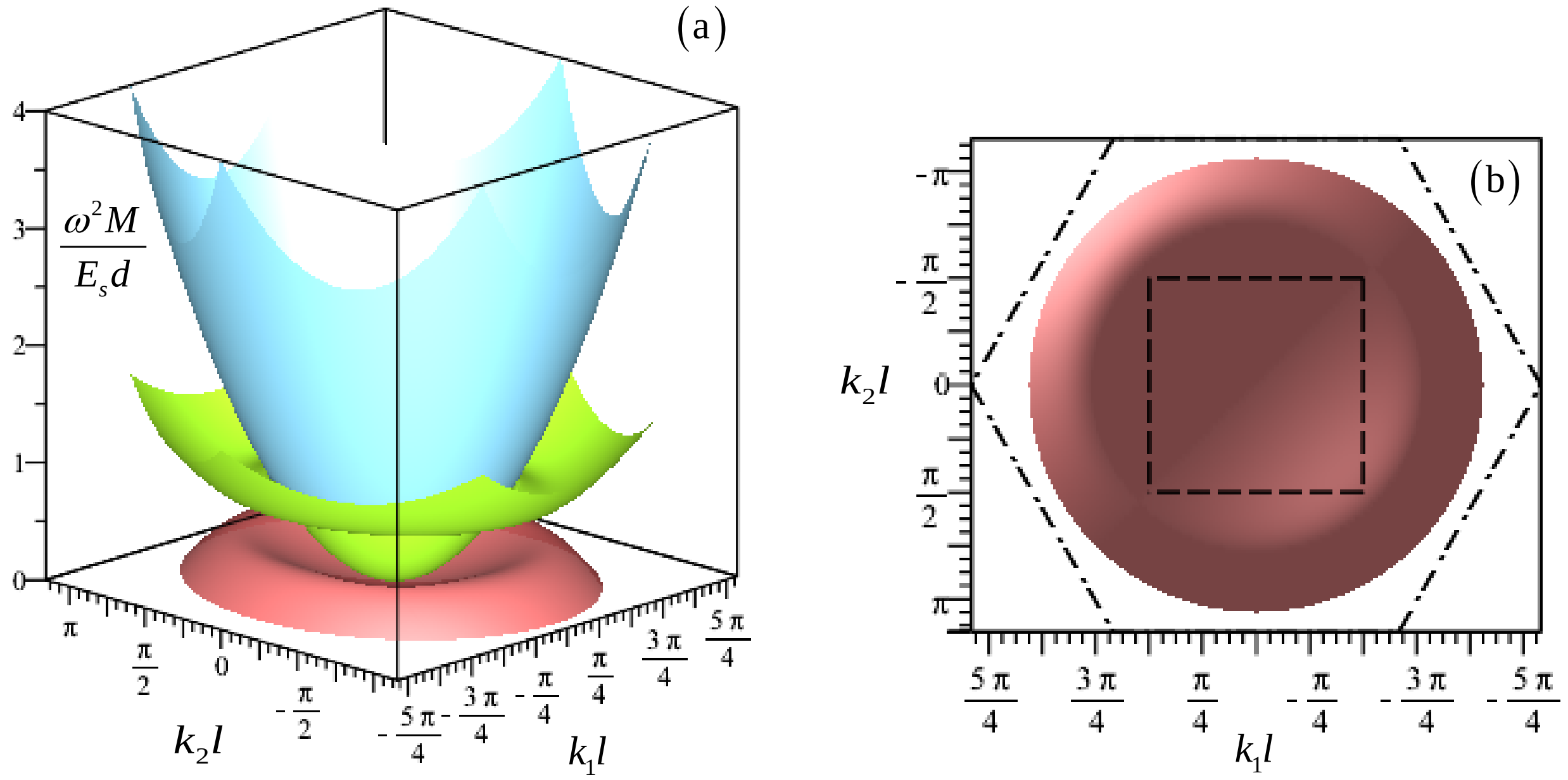

Figure 18: Square of the dispersive surfaces in the Brillouin zone by the micropolar continuum model for centrosymmetric lattice with $t/l = 1/5$, $J/Ml^2 = 1/50$, $\alpha = 1$. (a) Positive Floquet-Bloch spectrum; (b) Domain of positivity of the first acoustic surface and square subdomain of good accuracy of the model.

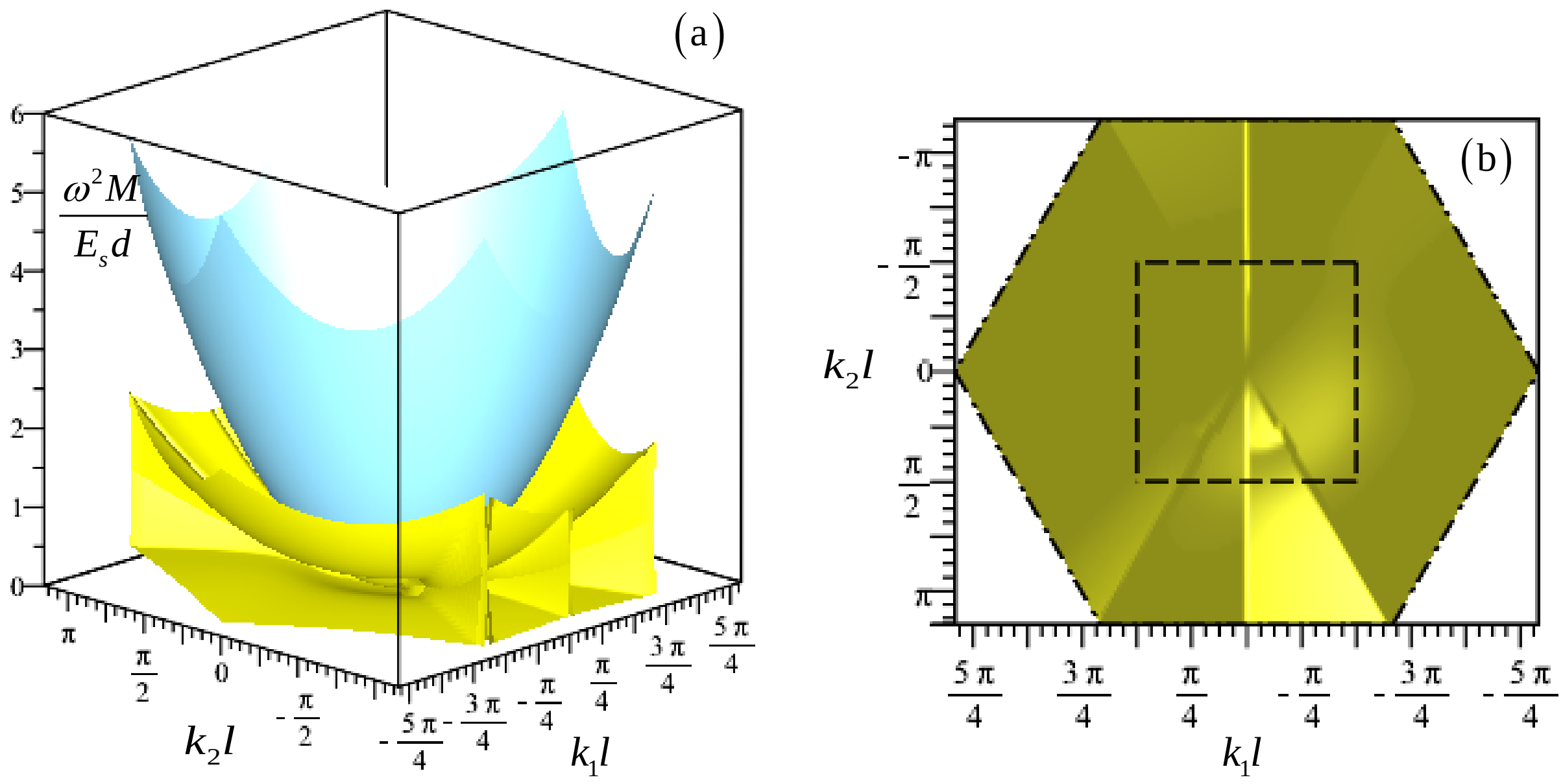

Figure 19: Square of the dispersive surfaces in the Brillouin zone by the micropolar continuum model for non-centrosymmetric lattice with $t/l = 1/5$, $J/Ml^2 = 1/8$, $\alpha = 2$. (a) Positive Floquet-Bloch spectrum; (b) Domain of positivity of the first acoustic surface and square subdomain of good accuracy of the model.

## 6. Conclusions

The in-plane acoustic behavior of non-centrosymmetric lattices having nodes endowed with mass and gyroscopic inertia and connected by massless ligaments with asymmetric elastic properties has been analysed through a discrete model and a continuum micropolar model. In the first case the propagation of harmonic waves and the dispersion functions have been obtained by the discrete Floquet-Bloch approach. In case of non-centrosymmetric lattice, the resulting eigenvector problem is ruled by a full matrix and two acoustic branches and an optical branch are obtained in the frequency spectrum. Moreover, it is shown that in general the optical branch departs from a critical point with vanishing group velocity and is decreasing for increasing wave vector from the long wave limit. The micropolar continuum model, useful to approximate the discrete model, has been derived through a continualization of the discrete equations of motion that is based on an approximation of the generalized displacements of the cells through an upscaling relation based on a second-order Taylor expansion of the generalized macro-displacement field. The equations of motion of the non-centrosymmetric lattice have been obtained together with the homogenized constitutive equations involving also the coupling third-order elasticity tensors of the micropolar continuum. It is worth noting that the second order elasticity tensor coupling curvatures and micro-couples turns out to be negative defined also in the general case of non-centrosymmetric lattice. This outcome has been obtained also through an extended Hamiltonian derivation and the Hill-Mandel macro homogeneity condition, based on a proper treatment of the potential elastic energy in terms of a second order expansion of the generalized displacement field following a suggestion by Bazant and Christensen, 1972, in modelling rectangular frames. A completely different outcome is obtained if a first order expansion of the rotational field is considered by applying the Hill-Mandel condition when the second order tensor turns out to be positive defined. In this regard, it is worth noting that the eigenvalue problem governing the harmonic propagation in the micropolar non-centrosymmetric continuum results in general characterized by a hermitian full matrix that is exact up to the second order in the wave vector.

Some examples have been analysed concerning square and equilateral triangular lattices and their acoustic properties have been obtained from both the exact Lagrangian model (within the assumed hypotheses) and the micropolar approximate model. For each lattice, the equations of motions have been given together with the constitutive parameters. The analysis of the influence of the model parameters on the acoustic behavior has shown that the non-centrosymmetry topology of the lattice may contribute to obtain low frequency band gaps.

Moreover, as occurs in the Lagrangian model, the optical dispersion branch is observed to be decreasing for increasing the norm of the wave vector from the long wavelength limit. On the contrary, it may be easily verified that if the elastic second order positive defined tensor is assumed, which is derived by a first order expansion of the rotation field, the optical branch turns out to be approximated by the equivalent micropolar continuum with a lower accuracy. This result, which is valid for the case of the propagation of harmonic waves, contributes to clarifying the problem highlighted by Kumar and McDowell, 2004, and thereafter by Liu *et al.* 2012, regarding the choice of the second order elastic tensor provided from the micropolar homogenization. In consideration of the negative definiteness of the second order elastic tensor of the micropolar model, the hyperbolicity of the equation of motion has been investigated by considering the Legendre–Hadamard ellipticity conditions requiring real values for the wave velocity. Since unconditional hyperbolicity cannot be ensured, some meaningful cases have been successfully considered by controlling the positivity of the square of the dispersion functions in the Brillouin zone. Finally, it is believed that the ellipticity of the field equations of the equilibrium problem could be recovered in a homogeneous continuum obtained through a higher order generalized micropolar homogenization involving higher order gradients of the macro-displacement field.